\documentclass[twocolumn]{aastex631} 
\usepackage{amsmath}
\usepackage{mathtools}

\graphicspath{{./}{figures/}}

\received{August 1, 2022}
\revised{October 31, 2022}
\accepted{November 3, 2022}

\shorttitle{Mission to an Interstellar Object}
\shortauthors{Siraj et al.}

\begin{document}

\title{Physical Considerations for an Intercept Mission to a 1I/'Oumuamua-like Interstellar Object}

\author{Amir Siraj}
\affiliation{Department of Astronomy, Harvard University, 60 Garden Street, Cambridge, MA 02138, USA}
\author{Abraham Loeb}
\affiliation{Department of Astronomy, Harvard University, 60 Garden Street, Cambridge, MA 02138, USA}
\author{Amaya Moro-Mart\'in}
\affiliation{Space Telescope Science Institute, 3700 San Martin Drive, Baltimore, MD 21218, USA}
\affiliation{William H. Miller III Department of Physics \& Astronomy, Johns Hopkins University, 3400 N. Charles Street, Baltimore, MD 21218}
\author{Mark Elowitz}
\affiliation{Harvard-Smithsonian Center for Astrophysics, 60 Garden Street, Cambridge, MA 02138}
\author{Abigail White}
\affiliation{Harvard-Smithsonian Center for Astrophysics, 60 Garden Street, Cambridge, MA 02138}
\author{Wesley A. Watters}
\affiliation{Whitin Observatory, Department of Astronomy, Wellesley College, Wellesley, Massachusetts, USA,}
\author{Gary J. Melnick}
\affiliation{Harvard-Smithsonian Center for Astrophysics, 60 Garden Street, Cambridge, MA 02138}
\author{Richard Cloete}
\affiliation{Department of Astronomy, Harvard University, 60 Garden Street, Cambridge, MA 02138, USA}
\author{Jonathan Grindlay}
\affiliation{Department of Astronomy, Harvard University, 60 Garden Street, Cambridge, MA 02138, USA}
\author{Frank Laukien}
\affiliation{Harvard-Smithsonian Center for Astrophysics, 60 Garden Street, Cambridge, MA 02138}

\begin{abstract}
In this paper, we review some of the extant literature on the study of interstellar objects (ISOs). With the forthcoming Vera C. Rubin Telescope and Legacy Survey of Space and Time (LSST), we find that $0.38 - 84$ `Oumuamua-like interstellar objects are expected to be detected in the next 10 years, with 95\% confidence. The feasibility of a rendezvous trajectory has been demonstrated in previous work. In this paper, we investigate the requirements for a rendezvous mission with the primary objective of producing a resolved image of an interstellar object. We outline the rendezvous distances necessary as a function of resolution elements and object size. We expand upon current population synthesis models to account for the size dependency on the detection rates for reachable interstellar objects. We assess the trade-off between object diameter and occurrence rate, and conclude that objects with the size range between a third of the size and the size of `Oumuamua will be optimal targets for an imaging rendezvous. We also discuss expectations for surface properties and spectral features of interstellar objects, as well as the benefits of various spacecraft storage locations.
\end{abstract}


\section{Introduction} \label{sec:intro}

Observations and numerical simulations indicate that the young solar system was very efficient at ejecting planetesimals into interstellar space as a result of planet migration and resonance sweeping \citep{Tsiganis2005,Morbidelli2005}. Given how ubiquitous planetary systems are, and that these planetesimal-clearing processes operate under a wide range of planetary architectures \citep{Raymond2018b}, it has been predicted that the Galaxy contains a large population of planetesimals ejected from planetary systems in the making. Another source of interstellar planetesimals are the clouds of comets loosely bound to  stars, like the solar system Oort cloud, from where comets can be easily ejected into interstellar space as the result of perturbations by passing stars, the galactic tide, or stellar mass loss at the end of a star's life. Interestingly, when these ejected interstellar planetesimals cross star- and planet-formation environments, they might get trapped with the help of gas drag and act as “seeds” of planet formation, helping overcome the meter-size barrier \citep{Pfalzner2019, Pfalzner2021, Moro2022}.

It was expected that, eventually, one of these ejected interstellar planetesimals would cross the orbit of the solar system and become an interstellar comet. Based on the expected contribution to the population of interstellar planetesimals from planet-forming systems, \citet{Moro2009} and \citet{Engelhardt2014} estimated that there were $n\sim5\times 10^{-5}\,$\textsc{au}$^{-3}$ objects with effective radius $r>0.5\,{\rm km}$ and $n<1.4\times 10^{-4}\,$\textsc{au}$^{-3}$ in local interstellar space. Therefore, the discovery of the first interstellar object, 1I/2017 U1 (`Oumuamua) \citep{Meech2017}, was a welcome advancement in the field. However, even though its detection was expected, its properties were not.

`Oumuamua was remarkably small (55–130 m) \citep{Jewitt2017,  Meech2017, Fraser2017, Drahus2018, Bolin2017} and could only be detected by the Panoramic Survey Telescope and Rapid Response System (Pan-STARRS) because it happened to pass very close to the Earth on its way out of the solar system. Only a few solar system objects of that size have been studied and they are all near-Earth objects, limiting our ability to make comparisons. Using the Pan-STARRS survey volume and time span (leading to a detection frequency), and assuming `Oumuamua represents a background population of interstellar objects, the number density of interstellar objects was inferred \citep{Trilling2017,Laughlin2017,Jewitt2017} and was found to be higher than expected in the context of a range of plausible origins \citep{Zwart2018, Raymond2017, Rafikov2018, Do2018, moro2018, moro2019a}, raising the question whether `Oumuamua was representative of an isotropic distribution of ejected objects. 

It is expected that the typical velocity of an interstellar object increases with time due to perturbations by passing stars, molecular clouds, the Milky Way spiral arms, and star clusters. However, `Oumuamua featured a low incoming velocity with respect to the local standard of rest \citep{mamajek2017}. This implies a relatively nearby source and a young $<40$ Myr age \citep{Fernandes2018,hallatt2020dynamics,Hsieh2021}. The latter is consistent with an excited rotational state \citep{Drahus2018, Fraser2017, Mashchenko2019}.  It terms of its color, it exhibited a reddened reflection spectrum consistent with iron-rich minerals and with space weathered surfaces and resembled small bodies in the solar system, including D-type asteroids, some Jupiter Trojans and trans-Neptunian objects, and comets \citep{Meech2017, Jewitt2017, Bannister2017, Knight2017, masiero2017spectrum, Bolin2018, Fitzsimmons2017, ISSI2019}. The lack of an ultra-reddened spectrum suggests `Oumuamua has not been exposed to space weathering for Gyr \citep{Gaidos2017a, Feng2018, Fitzsimmons2017}. 

Immediately apparent, `Oumuamua lacked any sign of a cometary tail in deep composite imaging (from dust or volatile outgassing), despite its small perihelion distance of 0.25 AU  \citep{Meech2017,Jewitt2017,Fitzsimmons2017, Trilling2018}. This was unexpected because models show that the majority of the objects ejected during planet formation originate beyond the snowline in their parent planetary systems and are therefore are expected to have a cometary nature. Most strikingly, in spite of the lack of apparent cometary activity, `Oumuamua exhibited a significant non-gravitational acceleration \citep{Micheli2018} that declined smoothly with distance from the Sun, showing no change in spin or sudden kicks, as routinely observed from jets on the surface of comets \citep{Rafikov2018}, and no apparent cut-off at the distance beyond which evaporation of water ice by sunlight is expected to stop. The Spitzer telescope data, sensitive to CO outgassing, ruled out the level of evaporation that would have been required to account for the observed non-gravitational acceleration if it was caused by the rocket effect from normal cometary outgassing amounting to about 10\% of its mass, \citep{Micheli2018}, unless the object chemical composition differed significantly from that of small bodies in the solar system \citep{fuglistaler2015solid,Trilling2018, ISSI2019, seligman2019acceleration, seligman2020,Levine2021_h2, Seligman2021, jackson20211i,desch20211i}. 

Also remarkable was `Oumuamua's drastic brightness variations, that combined with the phase angle indicated a fairly extreme $6:6:1$ geometry \citep{Meech2017, Knight2017, Bolin2017, McNeill2018, Belton2018, Drahus2018, seligman2019acceleration}, found to be disk-like at the ~90\% confidence level \citep{Mashchenko2019}. However, the shape remained uncertain because there were not enough observations sampling different phase angles and also because it is unknown to what degree the surface albedo variations contributed to the brightness variability.

With the detection of the second interstellar interloper, C/2019 2I/Borisov, it has become evident that the galaxy contains a massive population of interstellar objects. Unlike `Oumuamua, 2I/Borisov had a clear cometary nature. It exhibited a distinct dusty coma \citep{Jewitt2019b,Bolin2019,Fitzsimmons:2019,Ye:2019,McKay2020,Guzik:2020,Hui2020,Kim2020,Cremonese2020,yang2021} with traces of carbon and nitrogen based species \citep{Opitom:2019-borisov, Kareta:2019, Lin2020,Bannister2020,Xing2020,Aravind2021}, and it was particularly enriched in CO compared to H$_2$O, with an abundance ratio more than three times higher than in any comet in the inner solar system \citep{Bodewits2020, Cordiner2020}. This indicates that 2I/Borisov is an ejected ice-rich planetesimal that formed in the outermost region of its planetary system. 2I/Borisov's coma made the determination of the size of its nucleus challenging, estimated to be 200 m--500 m, without evidence of being as elongated as 1I/'Oumuamua \citep{Jewitt2020}. It was also found to exhibit an abnormally high polarization \citep{Bagnulo2021}. Interestingly, the comet produced atomic nickel vapor \citep{Guzik2021} and experienced a rotational bursting of one or more meter-size boulders \citep{Drahus2020ATel,Jewitt2020:BorisovBreakup, Jewitt2020ATel, Bolin2020ATel,Zhang2020ATel,Kim2020}. While 2I/Borisov exhibited non-gravitational acceleration similar to `Oumuamua, the astrometric positions were consistent with observed levels of outgassing  \citep{Hui2020,delafuente2020,Manzini2020}. 

The contrast between the expected properties of 2I/Borisov and the unexpected properties of `Oumuamua highlighted the elusive nature of the latter, leading to a variety of hypothesis regarding its provenance that contemplated objects that have never been seen before. These hypotheses can broadly be categorized into the postulated source of the anomalous acceleration and they all have theoretical and/or observational shortcomings \citep{Hoang2020,Levine2021,Siraj2022}.

If the acceleration was caused by cometary outgassing, as first proposed by \citet{Micheli2018}, which would be consistent with the observed brightness variations \citep{seligman2019acceleration}, this could be explained if the outgassing were dominated by H$_2$, \citep{fuglistaler2015solid,seligman2020,Levine2021_h2}, CO \citep{Seligman2021}, or N$_2$ \citep{jackson20211i,desch20211i}, implying `Oumuamua was like an "iceberg" that fragmented from a molecular cloud (if dominated by H$_2$ or CO) or chipped off the surface of a planet like Pluto (if dominated by N$_2$).

If the acceleration was caused by radiation pressure, as first discussed by \citet{Micheli2018}, this could be explained if the object had an ultra porous structure \citep{moro2019fractal,Sekanina2019b, luu2020oumuamua}
or an extremely thin geometry \citep{bialy2018radiation}. The former hypothesis implies that `Oumuamua was a cosmic ``dust bunny" with a bulk density about 100 times less than air. Numerical simulations indicate that such structures could form as the result of the collisional growth of icy dust particles beyond the snowline of a protoplanetary disk \citep{Okuzumi2012}; if this were its origin, it would make `Oumuamua a never seen before intermediate product of planet formation \citep{moro2019fractal}. Alternatively, it has been proposed that an ultra-porous structure grew from a collection of dust particles in the coma of an active comet that then escaped \citep{luu2020oumuamua}, or from the disintegration of an ``ordinary" km-sized extrasolar comet as it passed near the sun \citep{Sekanina2019}. The thin-geometry hypothesis implies that `Oumuamua could have been a new kind of natural astrophysical object or the remnant of an extraterrestrial civilization  \citep{Loeb2018a, Loeb2018b, Loeb2018c, bialy2018radiation, Loeb2021a, Loeb2021b}. In the context of the latter extraordinary hypothesis, it is of interest that in September 2020, another object was discovered by Pan-STARRS, sharing some of `Oumuamua's anomalous properties, such as exhibiting an excess acceleration but no cometary outgassing.. Given the astronomical name 2020 SO (MPEC 2020-S78:2020 SO), it was later confirmed to be the lost rocket booster from NASA’s failed 1966 Surveyor 2 mission to the Moon; its thin walls resulted in a large area-to-mass ratio that allowed the object to be pushed by radiation pressure, accounting for its non-gravitational acceleration.

It has been estimated that the Vera C. Rubin Observatory Legacy Survey of Space and Time (LSST) \citep{jones2009lsst,Ivezic2019}, whose ability to detect transient objects has been demonstrated \citep{solontoi2011comet,Veres2017,veres2017b,Jones2018}, will be able to detect 1-10 interstellar objects of `Oumuamua’s size or larger every year \citep{Engelhardt2014,Cook2016,Trilling2017}. This will shed light on their population and  possible origin. The detection of interstellar objects by LSST will also enable follow-up studies for detailed characterization by ground-based and spaced-based facilities, including by future intercept or rendezvous missions. An extraordinary opportunity has opened to study up-close planetesimals ejected from other planetary systems, as is surely the case of 2I/Borisov, and unusual interstellar objects like `Oumuamua, of particular interest for the Galileo Project. 

In this paper we discuss the characterization from close range of interstellar objects that, like `Oumuamua, don't have an unequivocally identified nature. Previous studies conclude that, based on the close proximity to Earth of the objects that are detectable by LSST, it would be feasible from an energetic standpoint to perform a space-based rendezvous mission \citep{Seligman2018,Hein2017,Meech2019whitepaper,Castillo-Rogez2019,Hibberd2020,Donitz2021,Meech2021,Hibberd2022,Moore2021whitepaper}; the European Space Agency \textit{Comet Interceptor} mission \citep{jones2019,Sanchez2021}  or the proposed NASA concept study \textit{Bridge} \citep{Moore2021} may provide us with this opportunity. Recently, \citet{Hoover2022} presented detailed dynamical simulations of the population of interstellar objects passing through the Solar System, and provided constraints on the distribution of the orbital properties for objects that will be detectable by LSST and reachable for an in-situ rendezvous. In this paper we build upon this work to assess the feasibility of an intercept mission to an interstellar object that, like `Oumuamua and 2020 SO (MPEC 2020-S78:2020 SO), exhibits a non-gravitational acceleration but no signs of cometary activity.

In Section \ref{sec:closeencounter}, we discuss the resolution achievable in the diffraction-limited and flux-limited regimes of a close encounter with an ISO, as well as the tradeoff between resolution and encounter timescale. In Section \ref{sec:lsstdetection}, we explore LSST's capability for detecting ISOs as a function of distance from the Earth and as a function of object size. In Section \ref{sec:missiondependence}, we then present predictions for the detection rate of reachable ISOs given a certain intercept $\Delta v$. Section \ref{sec:spectralfeatures} considers possible spectral features of ISOs detectable from close range, and Section \ref{sec:missionstorage} discusses mission storage and surveyor telescope locations. Finally, we summarize our conclusions in Section \ref{sec:conclusion}.


\section{Close Encounter With An ISO}
\label{sec:closeencounter}
\subsection{Diffraction-limited regime}
\label{subsec:diff}
\begin{figure}[]
\begin{center}
    \includegraphics[scale=0.4,angle=0]{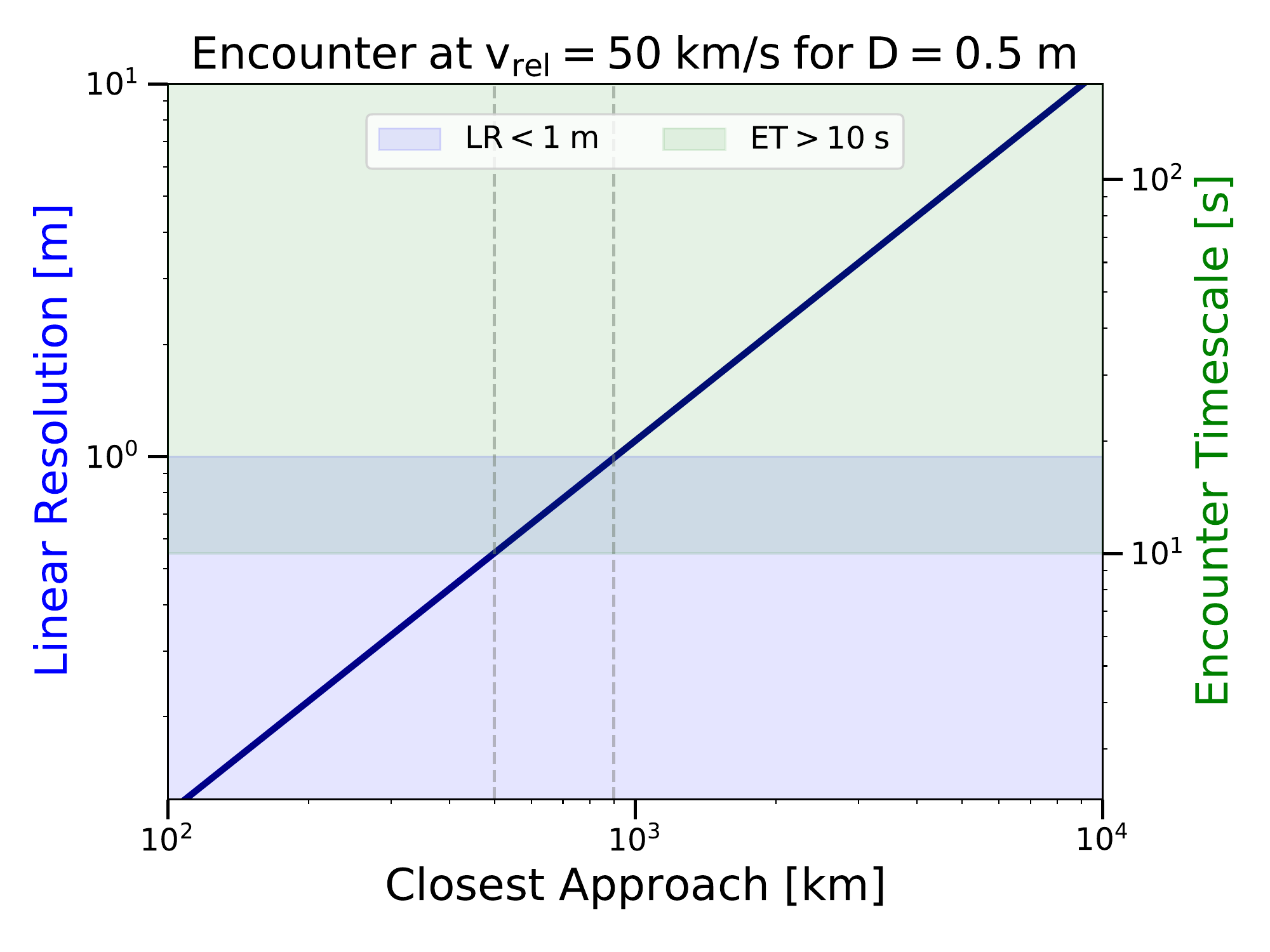}
    \caption{Linear resolution in the V-band as a function of closest approach distance to the interstellar object, with the corresponding encounter timescale on the secondary y-axis. Green shading denotes an encounter timescale of $> 10 \mathrm{\ s}$, while blue shading indicates a linear resolution of $< 1 \mathrm{\ m}$, for the fiducial parameters.}
    \label{fig:diffraction}
\end{center}
\end{figure}

The maximum number of linear resolution elements across an object in the diffraction limit is,

\begin{equation}
    N\sim D \,\bigg(\frac{d}{\lambda}\bigg)\, \bigg(\frac{1}{b}\bigg)\, ,
\end{equation}
where $\lambda$ is the wavelength that we observe with, $D$ is the aperture size of the telescope, $d$ is the effective spherical diameter of the object, and $b$ is the impact parameter or closest approach to the object. The encounter timescale is $t_{enc} = (b / v_{enc})$, and needs to be long enough for the spacecraft to capture measurements of the object. Figure \ref{fig:diffraction} shows diffraction-limited linear resolution and encounter timescale as a function of close approach to an ISO for a $0.5 \mathrm{\; m}$ telescope aperture and an encounter speed of $50 \mathrm{\; km \; s^{-1}}$, since this is a typical encounter speed between a gravitationally bound orbit and an unbound one near the Earth, $\sim \sqrt{(30 \mathrm{\; km \; s^{-1}})^2 + (40 \mathrm{\; km \; s^{-1}})^2}$. Given these parameters, an encounter timescale of $> 10 \mathrm{\ s}$ and a sub-$m$ linear resolution can be simultaneously achieved in the range $\sim (0.5 - 1) \times 10^3 \mathrm{\ km}$. This suggests that a close approach distance of several hundred km may be a `sweet spot' that balances linear resolution and encounter timescale for an ISO encounter mission. A close approach of $b \sim 10^3 \mathrm{\; km}$ with a $D \sim 50 \mathrm{\; cm}$ telescope would allow for linear resolution of $\sim 1 \mathrm{\; m}$, which would produce a $\sim 10^4$ pixel image of an object with a size of $\sim 100 \mathrm{\; m}$. In such a scenario, the spacecraft would have an encounter timescale of $\sim 20 \mathrm{s}$ to make the measurement. 

\subsection{Flux-limited regime}
At greater distances of closest approach, the amount of reflected sunlight received from the ISO limits the maximum number of resolution elements we can sample across the object. Here we calculate this maximum number of linear resolution elements as a function of closest approach for the flux-limited regime.

\begin{figure}[]
\begin{center}
    \includegraphics[scale=0.4,angle=0]{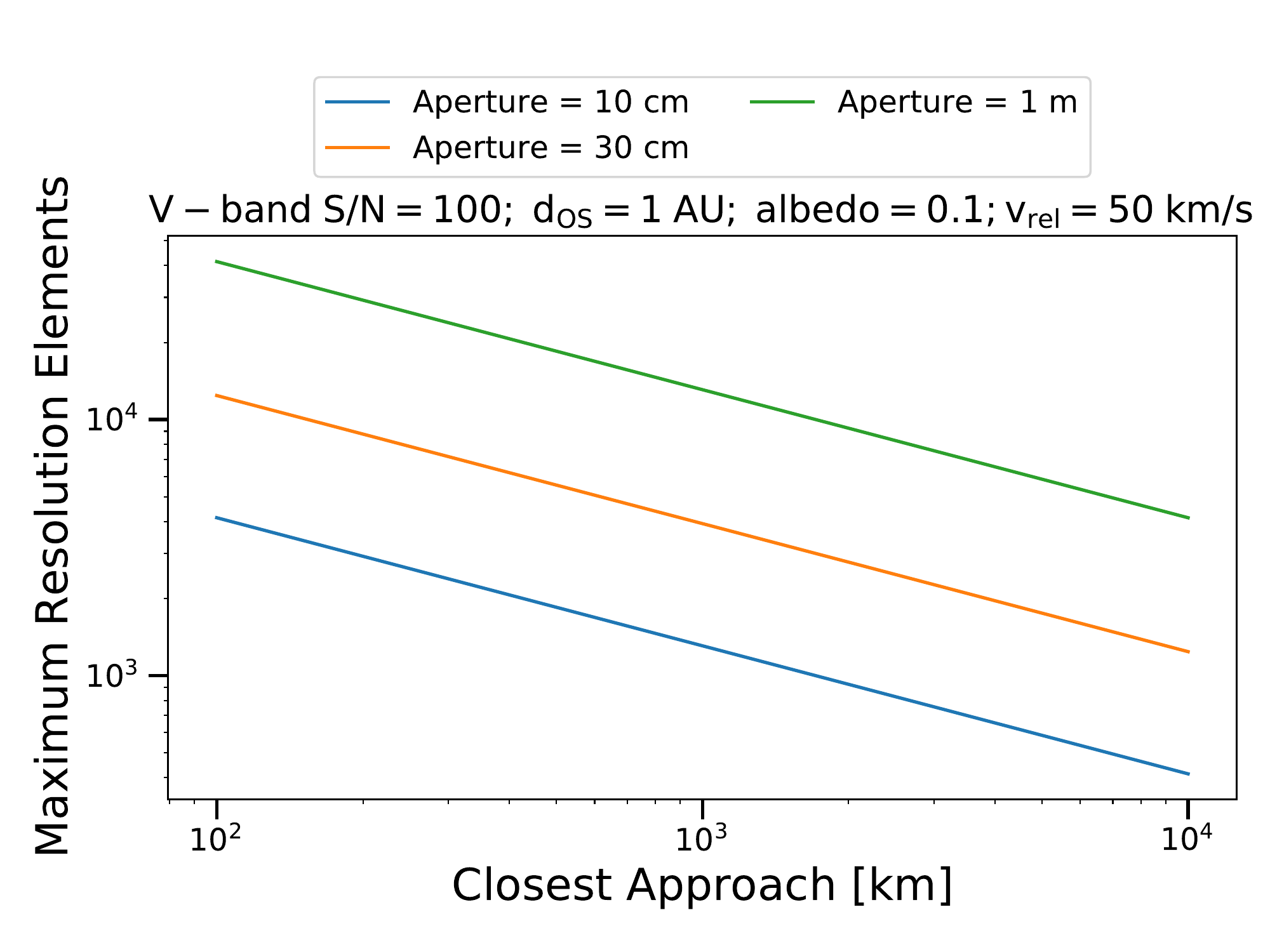}
    \caption{The flux-limited maximum number of resolution elements as a function of closest approach distance to an `Oumuamua-size object, given telescopes with apertures sizes of $10 \mathrm{\; cm}$, $30 \mathrm{\; cm}$, and $1 \mathrm{\; m}$, calculated using Equations \eqref{fluxelements} and \eqref{fluxelements2}.}
    \label{fig:resolutiondistance}
\end{center}
\end{figure}

The solar flux is $F_{\odot} = (L_{\odot} / 4 \pi d_{BS}^2)$, the photon energy is $E_{\gamma} = (hc / \lambda)$, the flux dilution factor at the telescope is $f_{FD} = (\pi D^2 / 4 \pi b^2) = (D^2 / 4 b^2)$, the object's cross-sectional area is $A = (\pi d^2 / 4)$, the encounter timescale is $t_{enc} = (b / v_{enc})$, the albedo is $\alpha$, the signal-to-noise ratio is $S/N$, and the fraction of the object's energy flux contained in the wavelength band centered at $\lambda$ is $f_{\lambda}$, where $L_{\odot}$ is the Solar luminosity, $d_{BS}$ is the distance of the object from the Sun, $h$ is the Planck constant, and $c$ is the speed of light. The total power at the telescope is $F_{\odot} f_{FD} f_{\lambda} \alpha A$, and therefore the number of photons reaching the telescope is $F_{\odot} f_{FD} f_{\lambda} \alpha A t_{enc} / E_{\gamma}$. This gives the constraint,
\begin{equation}
\label{fluxelements}
N < \left(\frac{S}{N}\right)^{-1} \sqrt{\frac{F_{\odot} f_{FD} f_{\lambda} \alpha A t_{enc}}{E_{\gamma}}} \; \; ,
\end{equation}
or equivalently, given the definitions above,
\begin{equation}
\label{fluxelements2}
N < \left(\frac{S}{N}\right)^{-1} \left( \frac{D \times d}{8 d_{OS}} \right) \sqrt{\frac{L_{\odot} \alpha f_{\lambda} \lambda}{b v_{enc} h c}} \; \; .
\end{equation}
To compute $f_{\lambda}$ and $E_{\gamma}$, we adopt the $V$ band, which is centered at $\lambda =551 \mathrm{\; nm}$ with a FWHM of $88 \mathrm{\; nm}$. 
The result is shown in Figure \ref{fig:resolutiondistance} for a signal-to-noise ratio of $S/N = 100$, a distance from the Sun of $d_{OS} = 1 \mathrm{\; AU}$, an albedo of $\alpha = 0.1$, and a relative encounter speed of $50 \mathrm{\; km \; s^{-1}}$. For these fiducial parameters telescope with an aperture of $10 \mathrm{\; cm}$ becomes flux-limited beyond $\sim 400$ resolution elements at a close-encounter distance of $\sim 10^4 \mathrm{\; km}$ from the object. This suggests that close encounters of the type mentioned in Section \ref{subsec:diff} at distances $\lesssim 10^3 \mathrm{\; km}$ are not flux-limited, and therefore diffraction-limited.

\section{LSST Detection Volume}
\label{sec:lsstdetection}

`Oumuamua’s absolute magnitude is $H = 22.4$ and we adopt a number density of $n \sim 0.1 \; \mathrm{AU^{-3}}$. LSST can see objects down to $m = 24$. We adopt a velocity dispersion at infinity of $\sigma \sim 40 \mathrm{\; km \; s^{-1}}$, since the velocity dispersion of interstellar objects is expected to be similar to that of local stars.

We use the phase functions described in Equations (8) – (10) of \cite{Hoover2022} to derive the apparent magnitude $m$ of an object with absolute magnitude $H$ at a distance from the Earth of $d_{BO}$, a distance from the Sun of $d_{BS}$, and with the distance from the Earth to the Sun being $d_{OS} = 1 \mathrm{\; AU}$. This allows us to compute the brightness of an object with a given absolute magnitude $H$ at every point in space.

The number density of ISOs is,
\begin{equation}
n \times \left[ 1 + \frac{2 G M_{\odot}}{d_{BS} \sigma^2} \right] \; \; ,
\end{equation} 
where $G$ is the gravitational constant and $M_{\odot}$ is the mass of the Sun. This allows us to compute number densities of ISOs at every point in space. 

\begin{figure}[]
\begin{center}
    \includegraphics[scale=0.5,angle=0]{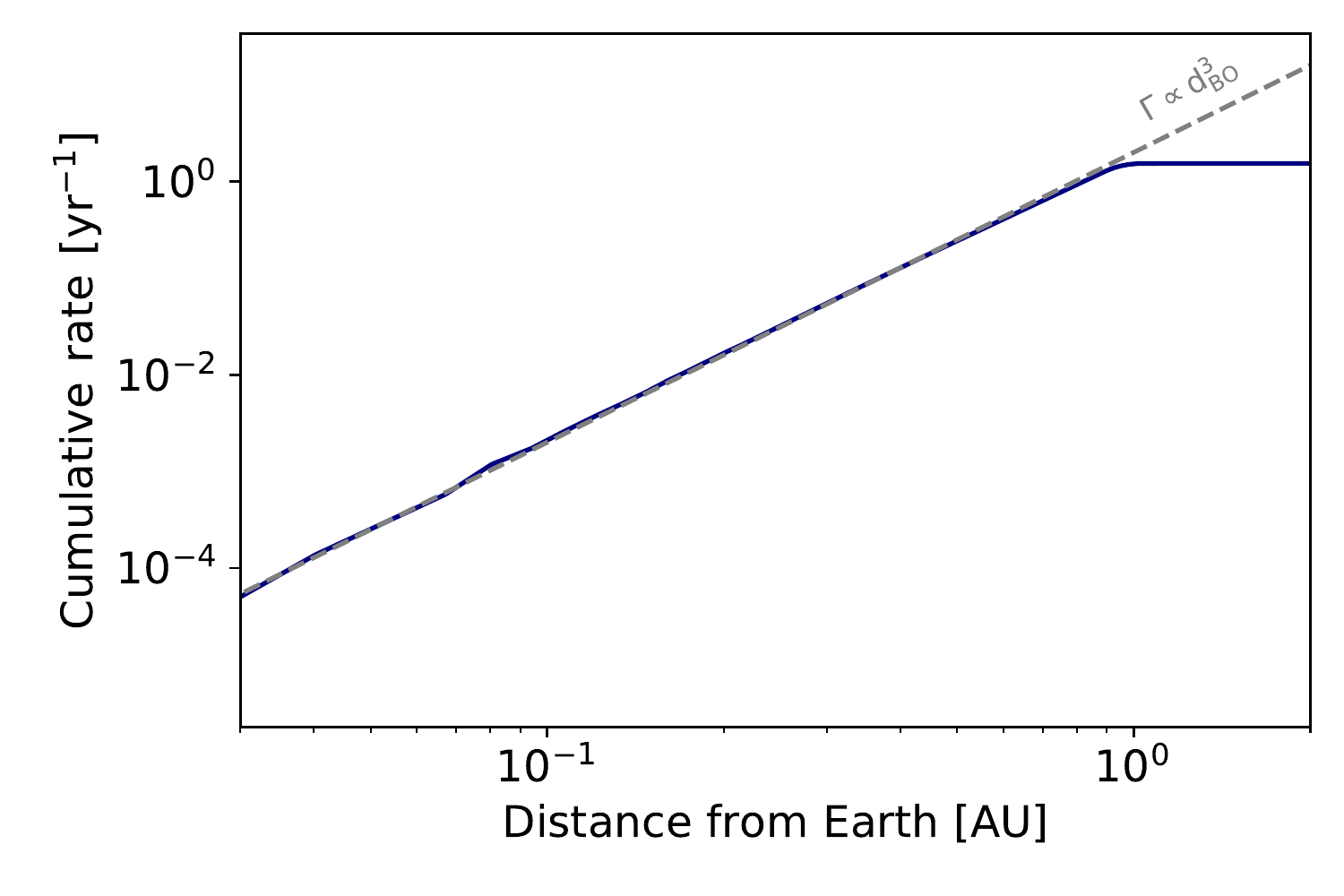}
    \caption{LSST detection rate for ISOs as a function of distance from the Earth. Power-law with a slope of 3 shown for reference, showing a strong match for $d_{BO} \lesssim \mathrm{1 \; AU}$.}
    \label{fig:detectionratescaling1}
\end{center}
\end{figure}

\begin{figure}[]
\begin{center}
    \includegraphics[scale=0.5,angle=0]{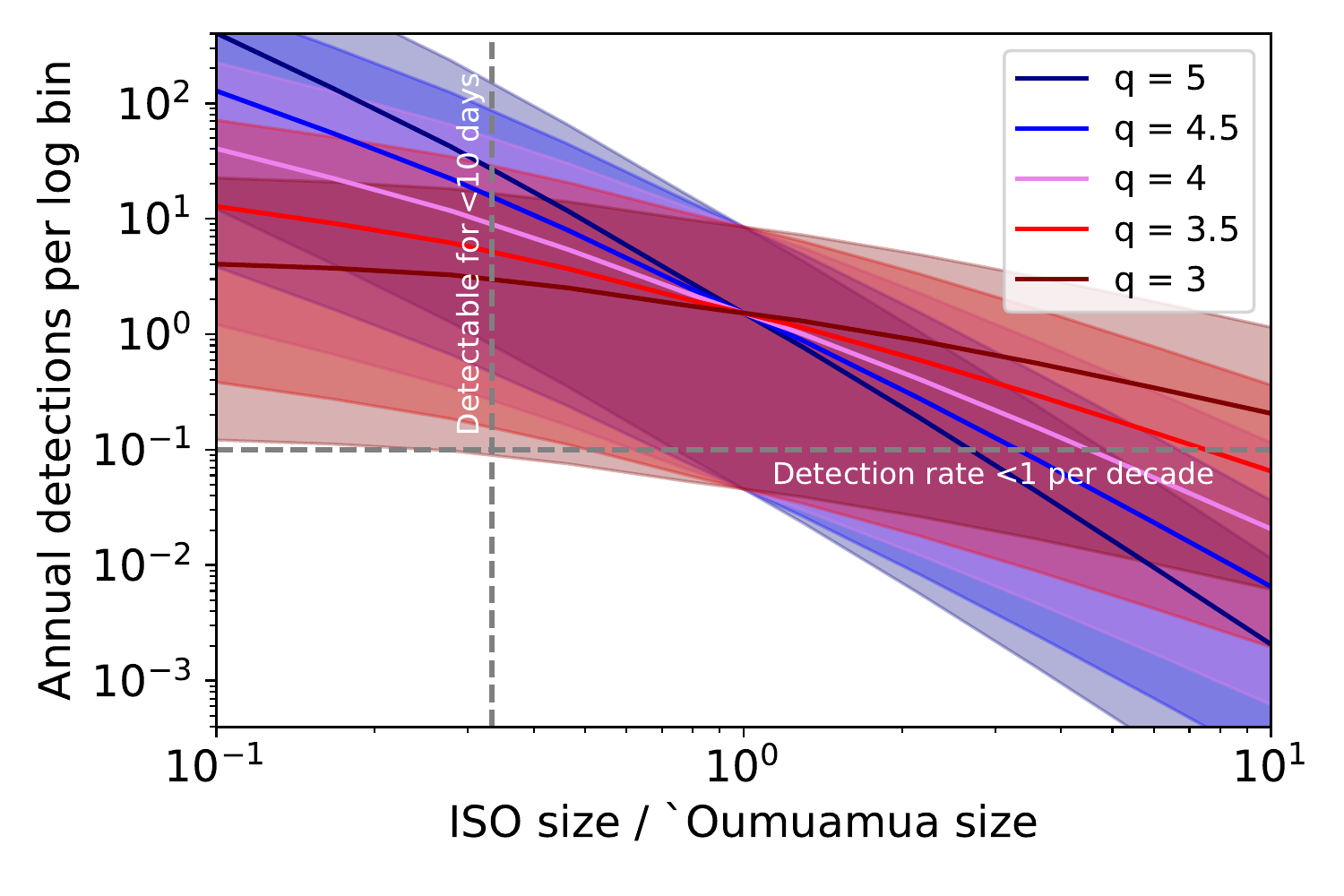}
    \caption{LSST detection rate for ISOs as a function of object size relative to `Oumuamua, adopting various values for the differential power-law size distribution index $q$ for the form $N(>d) \sim k d^{1-q}$. The shaded color bands correspond to 95\% confidence intervals given Poisson statistics for a single object. The dashed boundaries correspond to the thresholds of forbidden parameter space for LSST; crossing times $<$ 10 days and detection rates $<$ 1 per decade.}
    \label{fig:ratewithdistance}
\end{center}
\end{figure}

We then perform a cut for LSST’s sensitivity range of $m < 24$.
The crossing speed is,
\begin{equation}
v_{cross} \sim \sqrt{\frac{G M}{d_{BS}} + \sigma^2} \; \; ,
\end{equation}
and the crossing time is $t_{cross} = d_{cross} / v_{cross}$, where $d_{cross} \sim (4/3) V_{detect}^{1/3}$, and $V_{detect}$ is the detection volume. 
The crossing rate is then $\Gamma_{cross}  = t_{cross}^{-1}$, and the detection rate per unit volume is the product of the number density and the crossing rate. Integrating the detection rate per unit time per unit volume over the LSST detection volume then yields the overall detection rate. Fig.~\ref{fig:detectionratescaling1} shows that the detection rate for `Oumuamua-like objects follows a simple volumetric scaling up to $\sim 1 \mathrm{\; AU}$ away from the Earth and Fig.~\ref{fig:ratewithdistance} indicates the overall detection rate as a function of ISO size. The results are derived from the numerical simulation described above, by summing the ISOs as a function of radial distance from the Earth and computing the number of ISOs (of varying absolute magnitudes) within the LSST detection volume on the night side of the Earth, respectively. We consider size distributions of the form $N(>d) \sim k d^{1-q}$ with $2 \leq q \leq 5$ \citep{moro2018, moro2019a}, where $q$ is an index $k$ and a normalization factor. Note that objects smaller than $1/3$ the size of `Oumuamua (more than an order of magnitude lower in brightness) would be undetectable by LSST since the detection volume crossing time would be less than $0.1 \mathrm{\; AU \; } / (30 \mathrm{\; km \; s^{-1}} + \sigma^2) \sim 10 \mathrm{\; days}$, thereby prohibiting the 3 LSST detections required to constrain an orbit at the planned cadence of $\sim 100$ sky location visits per year.\footnote{\url{https://docushare.lsst.org/docushare/dsweb/Get/LPM-17}} ISOs with detection rates of $\lesssim 0.1 \mathrm{\; yr^{-1}}$ are also undetectable, given that LSST will operate for a decade.

We find that the number density of $n = 0.1 \mathrm{\; AU^{-3}}$ corresponds to 15 `Oumuamua-like objects over LSST’s 10 year lifespan, with a 95\% Poisson confidence interval yielding the range $0.38 - 84$. Poisson statistics imply that there is a $\sim 6 \%$ chance that LSST will not discover any `Oumuamua-like objects, and that there is a $\sim 20 \%$ chance that LSST will discover only one `Oumuamua-like object in the first three years of operation. \cite{2019arXiv190407224S} found that $q \sim 3.5$ may be implied by the detection of CNEOS 2014-01-08, which corresponds to $\sim 50$ objects with $H \sim 25$ over LSST’s lifetime, with a 95\% Poisson confidence interval of $1.3 – 280$.

\section{Mission Dependence on ISO Size}
\label{sec:missiondependence}

Table 2 from \citet{Hoover2022} suggests that the number of detectable ISOs with upper limits on the intercept $\Delta v$ ranging from $2 - 30 \mathrm{\; km \; s^{-1}}$ scales as $N(< \Delta v) \propto (\Delta v)^{k}$ where $k \sim 2.4$, and the fraction of all detectable objects that satisfy $\Delta v < 30 \mathrm{\; km \; s^{-1}}$ is $f \sim 0.4$, where $f \sim 1$ would represent all detectable objects. Furthermore, ranging from objects with sizes a factor of $4$ smaller than `Oumuamua to objects with similar sizes to `Oumuamua, $k$ and $f$ follow the respective empirical scalings, $k \propto d^{0.24}$ and $f \propto d^{-0.66}$. These empirical scalings serve as interpolation functions over $\Delta v$ parameter space; we do not extend them beyond the bounds enumerated in \citet{Hoover2022}. The power law indices presented here simply characterize the discrete results presented in Table 2 of \citet{Hoover2022} over the range $2 - 30 \mathrm{\; km \; s^{-1}}$ in a continuous fashion.

The ISO detection rate scales with volume ($d_{BO}^3$) but also with the differential size distribution, which is a power law with slope $-q$. Given some magnitude limit and albedo, the distance out to which we can see an object scales as the size of the object, $d_{BO} \propto d$, so we can express the dependence of the detection rate on $d$ as,

\begin{equation}
   \Gamma \propto d^{3-q}
\end{equation}

The LSST ISO detection rate is then,

\begin{equation}
    \label{eq:ratescaling}
    \Gamma \sim 1.5 \mathrm{\; yr^{-1}} f \left(\frac{\Delta v_{UL}}{\mathrm{30 \; km \; s^{-1}}} \right)^k \left(\frac{d}{d_O} \right)^{3 - q} \; \; ,
\end{equation}
where $f \sim 0.4 d^{-0.66}$ and $k \sim 2.4 d^{0.24}$, over the range $0.25 < (d / d_O) < 1$ and $2 \mathrm{\; km \; s^{-1}} < \Delta v_{UL} < 30 \mathrm{\; km \; s^{-1}}$, where $d_O$ is the diameter of `Oumuamua and  $\Delta v_{UL}$ is the upper limit on the required impulsive change in speed $\Delta v$. Similarly, 
\begin{equation}
    \label{eq:deltavscaling}
    \Delta v_{UL} \sim 30 \mathrm{\; km \; s^{-1}} \left(\frac{r}{r_O} \right)^{\frac{q - 3}{k}} \left( \frac{\Gamma}{1.5 f} \right)^{\frac{1}{k}} \; \; ,
\end{equation}
where $r$ is the effective radius of the object.

We find that, if the payload is launched as soon as an ISO is detected, there is an $\sim 85 \%$ chance of finding a suitable `Oumuamua-like target within LSST's 10-year life-span if the budget is $\Delta v_{UL} \sim 30 \mathrm{\; km \; s^{-1}}$, but this is reduced to a coin-flip if the budget is $\Delta v_{UL} \sim 15 \mathrm{\; km \; s^{-1}}$. For a budget of $\Delta v_{UL} \sim 10 \mathrm{\; km \; s^{-1}}$, the chance of finding a suitable `Oumuamua-like object in 10 years is only $\sim 33 \%$. The theoretical likelihood of success can be maximized by relaxing the size requirement to $(1/3)$ the size of `Oumuamua, yielding a $\gtrsim 90 \%$ chance for $\Delta v_{UL} \sim 30 \mathrm{\; km \; s^{-1}}$ , a $\gtrsim 75 \%$ chance for $\Delta v_{UL} \sim 15 \mathrm{\; km \; s^{-1}}$, and a $\gtrsim 60 \%$ chance for $\Delta v_{UL} \sim 10 \mathrm{\; km \; s^{-1}}$, but in reality the short timespan available to reach such an object may render the detection and launching procedures prohibitively difficult to design. The ideal size of an ISO for maximizing the likelihood of a suitable target within LSST's lifetime is therefore likely between $(1/3)$ and order unity relative to the size of `Oumuamua. It is also important to note that the linear resolution of the spacecraft's measurements is proportional to object size, so this must also be taken into account when optimizing mission parameters. The planned budget for ESA's \textit{Comet Interceptor} mission \citep{jones2019,Sanchez2021}, $\Delta v_{UL} \sim 15 \mathrm{\; km \; s^{-1}}$, gives a $\sim 0.02 \%$ and $\gtrsim 0.04 \%$ (but $\lesssim 15 \%$) likelihood of finding a suitable target within 10 years, for `Oumuamua-like objects and objects $(1/3)$ the size of `Oumuamua, respectively. 

\begin{figure}[]
\begin{center}
    \includegraphics[scale=0.5,angle=0]{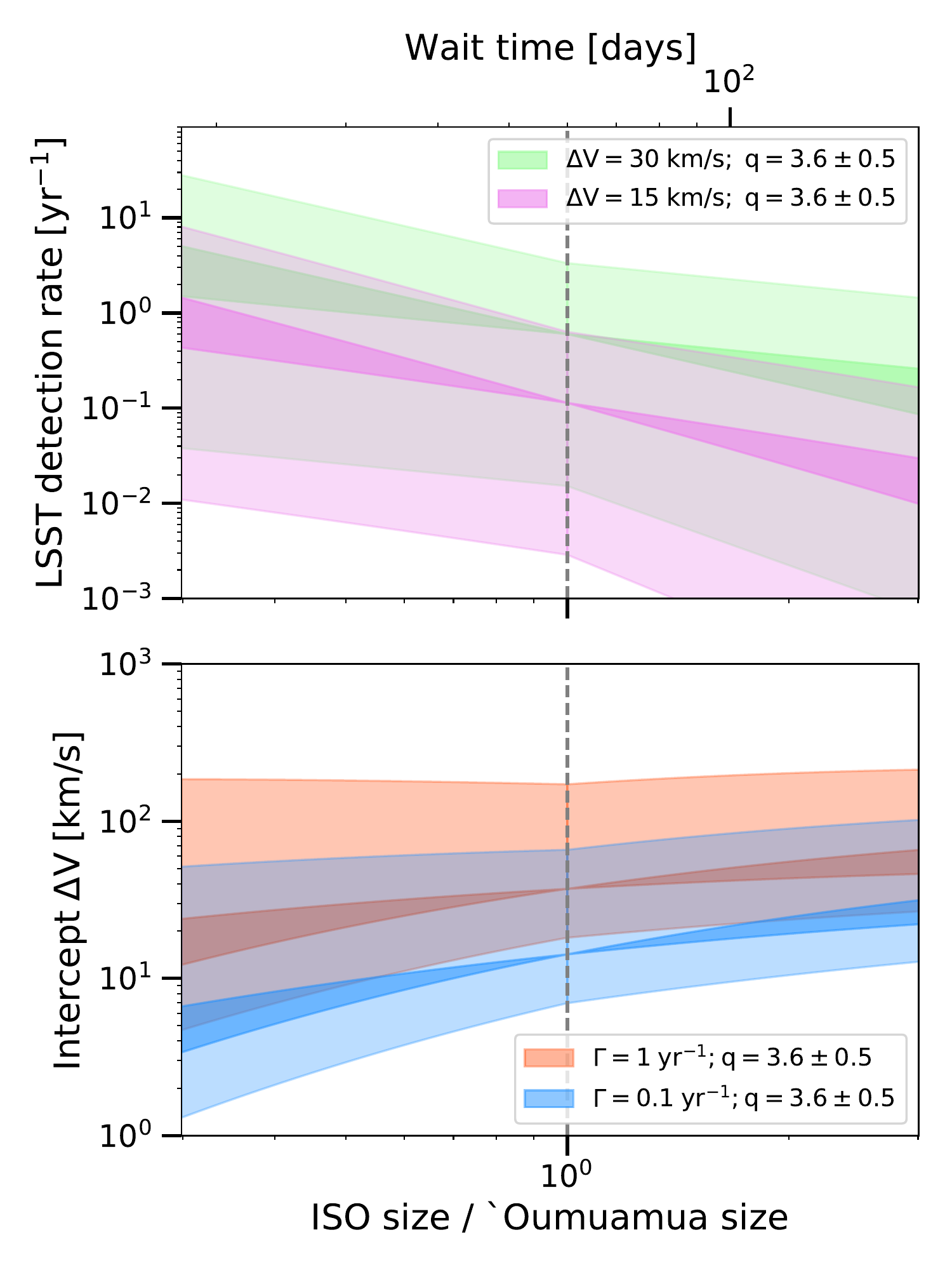}
    \caption{LSST detection rate and intercept $\Delta v$ as a function of interstellar object size or wait time. Dark regions correspond to $q = 3.6 \pm 0.5$ and light regions signify the 95\% Poisson confidence intervals for the light regions. The gray dashed line corresponds to an `Oumuamua-like object, and the plot shows results for an order of magnitude in size centered around `Oumuamua.}
    \label{fig:ISOsizecomparison}
\end{center}
\end{figure}

Wait time, $\tau$, defined as the time between the detection of a reachable object and its closest approach to the Earth, is roughly proportional to distance of detection $\tau \propto d_{BO}$, since the velocity dispersion of ISOs is expected to be comparable to or exceed the orbital speed of the Earth around the Sun, $\sigma \gtrsim \sqrt{2GM_{\odot}/R_{\odot}}$. Since $d_{BO} \propto r$ and the typical wait time for objects reachable with $\Delta v_{UL} \sim 20 \mathrm{\; km \; s^{-1}}$ is $\tau \sim 2$ months \citep{Hoover2022}, the scaling, $\tau \sim 2 \mathrm{\; months} \; (r / r_O)$, can be used to estimate the typical wait time to order of magnitude for objects similar in size to `Oumuamua. Note that the $\Delta v$ values, adopted from \citet{Hoover2022}, are based on wait time. Fig. \ref{fig:ISOsizecomparison} shows this approximation of wait time as a secondary x-axis for object size on the plots based on Equations \ref{eq:ratescaling} and \ref{eq:deltavscaling}, which illustrate the interplay between intercept $\Delta v$ and LSST detection rate. Furthermore, the wait time convention used here and in calculating delta-v is the most optimistic case possible because there is no delay between detection and mission launch. While we do not take into account any delay between detection and launch, we do take into account the time for the payload to reach the object.

\section{Surface properties and spectral features}
\label{sec:spectralfeatures}

Characterizing the surface properties of interstellar objects such as composition, reflectivity, and surface roughness, will be critical for each phase of this investigation.  First, this is vital for identifying interesting ISOs on their approach to the inner solar system, to decide whether to attempt a spacecraft intercept or rendezvous; and (2) in order to distinguish between hypotheses for the origin and provenance of these objects.

\begin{figure*}
\begin{center}
    \includegraphics[scale=0.4]{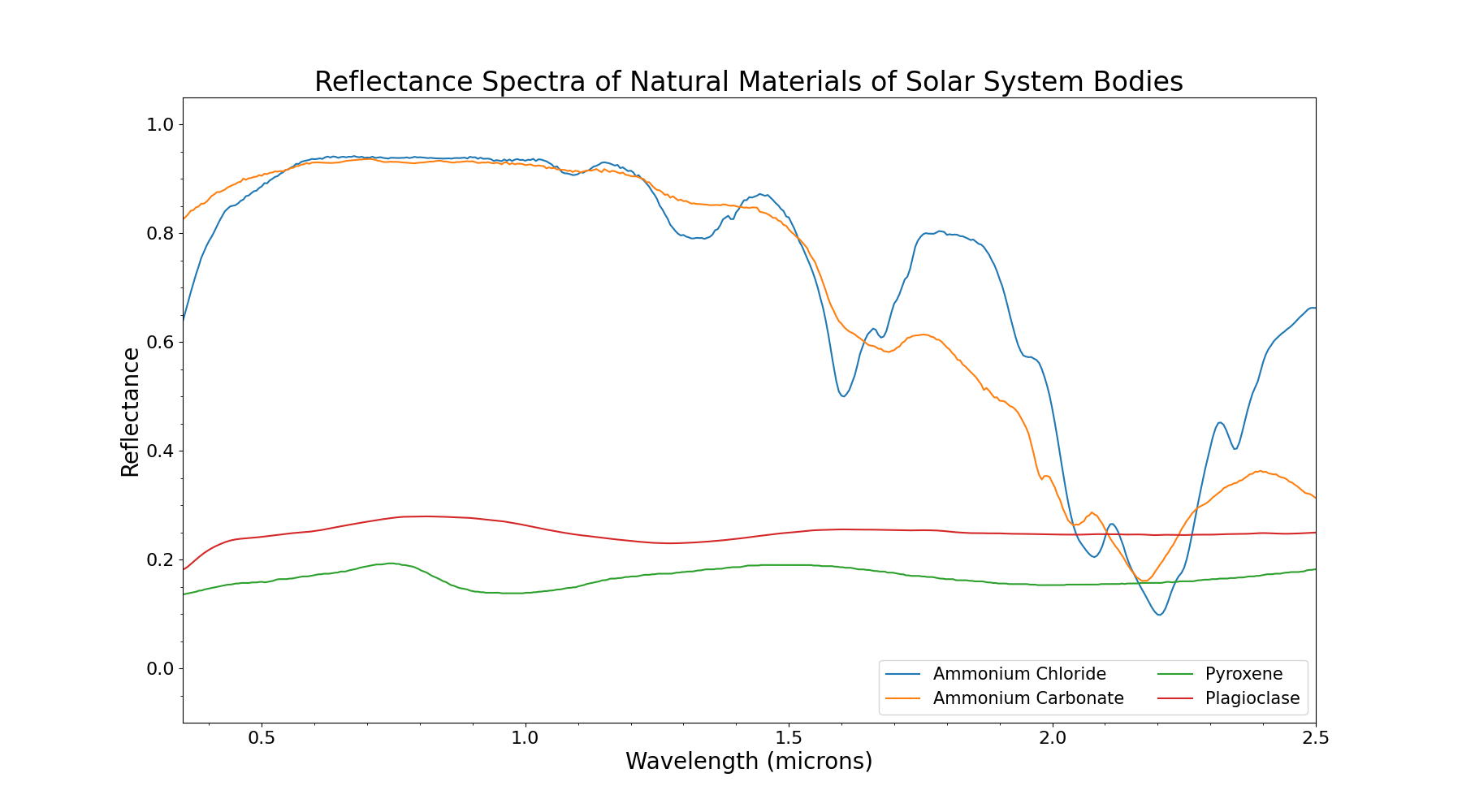}
\end{center}
\caption{Reflectance spectra measured in the laboratory over the UV to shortwave-IR spectral range ($\sim$0.4 to 2.5 microns). Each natural material shows a distinctive spectral signature that can be used for its detection and/or identification. (Sources: \cite{2014JGRE..119.1403V}, MRO CRISM Type Spectra Library, 2015, https://crismtypespectra.rsl.wustl.edu, USGS Spectral Library Version 7, 2017, https://doi.org/10.3133/ds1035)}
\label{fig:example_spectra}
\end{figure*}

\begin{figure*}
\begin{center}
    \includegraphics[scale=0.4]{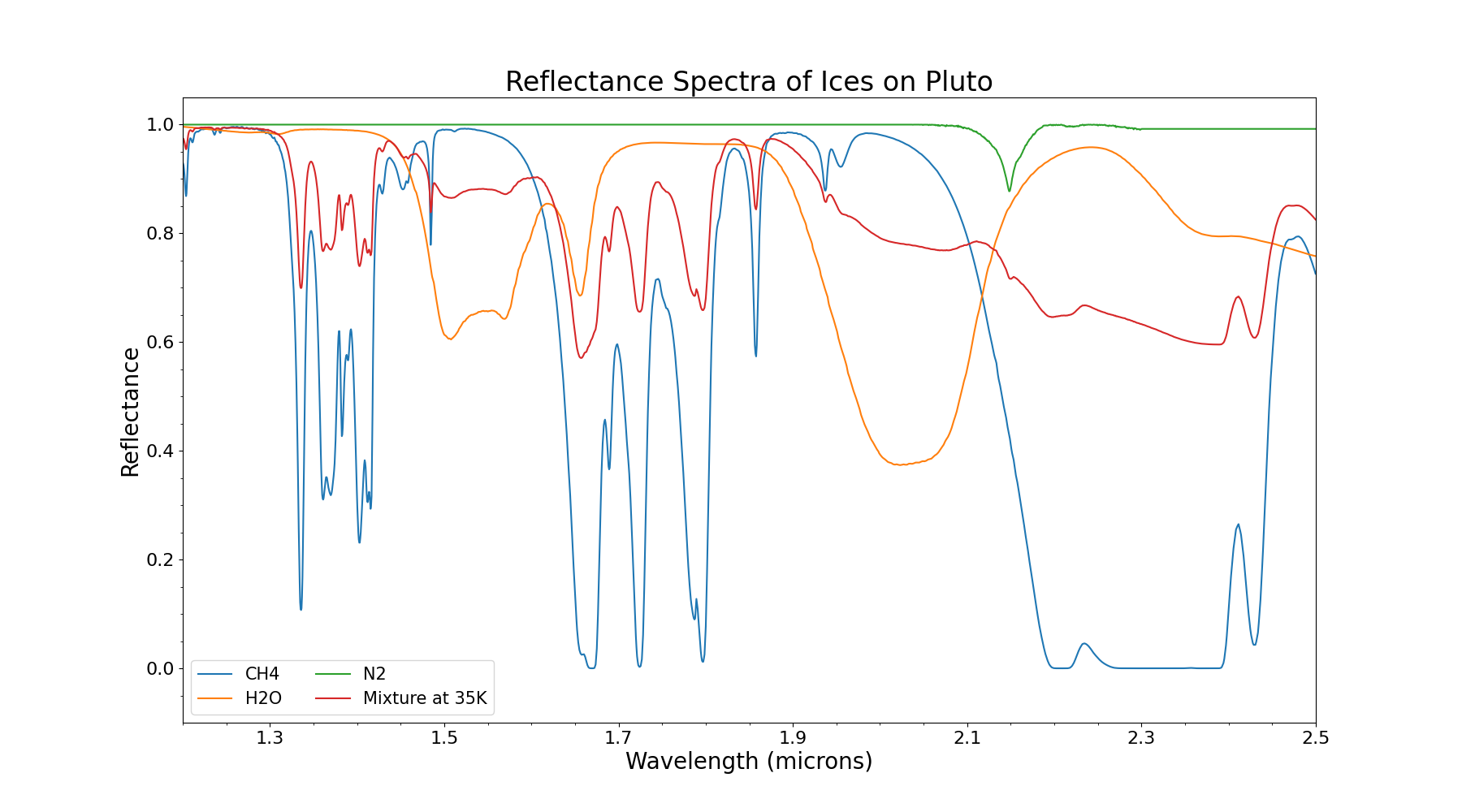}
\end{center}

\caption{Reflectance spectra of methane, water, and nitrogen ices, and a simulated ice mixture of 60 percent methane, 20 percent water, and 20 percent nitrogen at 35 Kelvin. Spectra range from 1.2 to 2.5 microns in the UV to shortwave-IR spectral range. Each ice shows a distinctive spectral signature that can be used for its detection and/or identification. (Source: NASA Planetary Spectrum Generator, 2022, https://psg.gsfc.nasa.gov)}
\label{fig:ice_spectra}
\end{figure*}

Possibly the most valuable measurement derives from reflectance and emission spectroscopy, which can provide a wealth of information on the mineralogical composition of planetesimals. Planetary space probes typically use imaging spectrometers to provide spatially resolved spectral maps of planetesimal surfaces during flyby and orbital survey missions. These spectrometers provide images, where each pixel represents both spatial and spectral information. The variation in mineralogical/chemical composition of the upper surface layers can inform geological maps to provide a geochemical survey of the object under study. To identify and quantify the abundance of a particular mineral within the surface layers, observed spectra are compared with reference spectra measured in laboratory experiments. The locations, depths, and absorption band areas of the observed spectral data can be compared to a library of reference spectra of varying band depths and areas to provide information about the minerals on the surface of a planetesimal (e.g., Pyroxene, Spinel, Olivine, Iron-Nickel alloy, etc.), including their modal abundances and grain size.  Example reflectance spectra of potential relevance to distinguishing between hypotheses for the provenance of `Oumuamua-like objects are shown in Figures \ref{fig:example_spectra}-\ref{fig:artificial_spectra}.

\subsection{Rocky materials}

Passive reflectance spectra over the wavelength range 0.4 to 2.5 $\mu$m can be used to discriminate different minerals based on the locations of absorption band minima. A common class of mineral found in asteroids is silicates, identified by the presence of broad absorption bands near 1 and 2 $\mu$m. Carbonaceous type asteroids show hydroxyl (-OH) near 2.7 $\mu$m, the presence of N-H at 3.05 $\mu$m, carbonate (C-H) near 3.4 $\mu$m, and possibly carbonate at 3.95 $\mu$m with absorption near 3.4 $\mu$m also noted. In general, the specific mineral detected using reflectance spectroscopy depends on electronic and vibrational transitions within the mineral crystals or molecules. Both ionic (e.g. Fe$^{2+}$, Fe$^{3+}$, etc.) and molecular (e.g. H$_2$O, CO$_3$, OH, etc.)  species are found in spectra of asteroids, each producing absorption features at different wavelengths. For the case of a mixture of different minerals on an asteroid's surface, various advanced spectral analysis techniques exist, including linear mixing, matched filter combined with partial spectral mixing (e.g. mixture-tuned matched filter), and machine learning algorithms. The accuracy of the identification of the minerals depends strongly on the quality/accuracy of the reference spectral library used in the analysis. Example spectra relevant to rocky surfaces are shown in figure \ref{fig:example_spectra}.

Laboratory experiments can also produce reference spectra that have been modified by various space weathering effects. This is important in performing a geochemical analysis of asteroids, since the effects of space weathering (e.g. irradiation by high-energy particles and micrometeoroid bombardment over geologically long periods of time) can alter the overall slope of the observed spectrum. Space weathering effects are more important for silicate type asteroids, which tend to show a darkened and reddened reflectance spectrum over the wavelength range 0.2 to 2.7 $\mu$m \citep{hapke2001space}. In laboratory experiments, the effects of cosmic rays and micrometeorites on asteroid surfaces, can be simulated using nanopulse lasers on various mineralogical samples. The lab experiments show that asteroid spectra can be darkened, reddened, and their absorption bands dampened by space weathering effects \citep{sasaki2001production,brunetto2006modeling}. It is therefore important to include the results of these laboratory experiments when analyzing the spectra of asteroids and interstellar objects. Furthermore, the addition of endmember spectra in mixtures of spectra can lead to a degeneracy (i.e. an ambiguity caused by more than one mineral having similar spectral profiles), and should be emphasized during the analysis reporting phase. Advanced techniques such as derivative spectroscopy can prove useful in accentuating weak or blended spectral features, thus mitigating to a certain degree the problem of spectral ambiguities encountered in the analysis of asteroid and interstellar object spectral data.

In the case of interstellar objects, spectroscopy can provide data that could be used to characterize the formation processes of ``exo-planetesimals'' or planetesimals that formed around other star systems. The spectral data could also be used to study the effects of long-term exposure of these objects due to their trajectories through interstellar medium. For example, the lack of surface ice as determined by reflective and infrared spectroscopy studies of ISOs could imply the possibility of an insulating or protective mantle produced by the object's exposure to cosmic rays over its long interstellar journey towards our solar system. For objects in the outer solar system with possible surface ice (e.g., Centaurs, KBOs, ISOs, etc.), spectroscopy beyond 2.5 $\mu$m (both MWIR and LWIR) can be used to detect the presence of phyllosilicates, organics and different frozen volatiles (e.g. water, ammonia, methane, etc), as well as the degree of hydration.

\subsection{Ices}

For KBOs in the outer solar system, where temperature are low enough, the dominant ice species seen in spectra include CO and CO$_2$, with minor amounts of reduced carbon (mostly CH$_4$, N$_2$, CH$_3$OH, H$_2$CO, and H$_2$S).  Spectral measurements of the icy surface of Pluto reveal the presence of methane (CH$_4$), nitrogen (N$_2$), ammonia (NH$_3$), and water ice (H$_2$O), with minor abundances of organic materials. Spectra of methane, nitrogen, and water ice are featured in Figure \ref{fig:ice_spectra}.

Water, carbon monoxide, carbon dioxide, and methane ice have characteristic spectral features in the 1-20 um range. In a 1997 study, Robert H. Brown and Dale P. Cruikshank collected laboratory reflectance spectra ranging from 1 to 5 $\mu$m of these ices. CO has strong absorption at 4.67 $\mu$m, which is Gaussian band centered, and a weak absorption of similar shape at 2.35 $\mu$m. The 2.35 $\mu$m band is a result of the first overtone of the fundamental transition of CO. For CO$_2$, there are several combination absorption bands in the 1.9-2.9 $\mu$m spectral region, more specifically two sets of combination absorption bands centered at about 2.0 and 2.7 $\mu$m. Additionally, strong absorption is exhibited at 4.3 $\mu$m. H$_2$O has characteristic absorption features at 1.5, 2.0, and 4.6 $\mu$m. 

CH$_4$ has four fundamental modes: 3.31 and 7.66 $\mu$m (infrared active) as well as 3.43 $\mu$m and 6.55 $\mu$m. \cite{1997A&A...317..929B} report that the infrared active fundamental modes (3.32 $\mu$m and 7.68 $\mu$m) are due to C-H stretching and deformation modes, $\nu_3$ and $\nu_4$. Note that the stretching mode at 3.32 $\mu$m can be entangled with the deep absorption band of water at 3.07 $\mu$m, complicating the ability to detect methane ice via spectra \citep{1997A&A...317..929B}. \citep{2018MNRAS.476.5432M} note eight absorption bands for methane ice. Two are attributed to v$_3$ and v$_4$ fundamental modes (3.32 and 7.69 $\mu$m), another two to 3v$_4$ and 2v$_4$ overtone modes (2.6 and 3.86 $\mu$m), and four to v$_2$+v$_3$, v$_3$+v$_4$, v$_1$+v$_4$, and v$_2$+v$_4$ combination modes (2.21, 2.33, 2.38, and 3.55 $\mu$m respectively).

Additionally, \cite{lynch2005infrared} report four features dominating the spectrum of water ice from 2-20 $\mu$m. The first is absorption at 3.16 $\mu$m due to O-H-O symmetric stretch vibration. Features at 6.1 $\mu$m and 4.6 $\mu$m are both attributed to the $\nu_2$ bending mode and a combination of $\nu_2$ and a weak libration mode, respectively. Another absorption feature highlighted by \cite{lynch2005infrared} is at 3 $\mu$m and is attributed to OH stretching. There is also a broad absorption centered at 13 $\mu$m due to a lattice absorption. Another laboratory study by \cite{1995A&A...296..810G} observed infrared band strengths of water, carbon monoxide, and carbon dioxide ices. H$_2$O band positions are reported as 3.045 $\mu$m (O-H stretch), 6.024 $\mu$m (O-H bend), and 13.16 $\mu$m (libration). Stretching of $^{12}$C$\equiv$O and $^{13}$C$\equiv$O are responsible for the 4.675 $\mu$m and 4.780 $\mu$m bands, respectively, observed for CO. As for CO$_2$, there are many band positions resulting from $\nu_3$, $\nu_2$, and combinations of $\nu_1$+$\nu_3$, and 2$\nu_2$+$\nu_3$. Similarly to CO, there are two bands from $^{12}$C$\equiv$O and $^{13}$C$\equiv$O stretch, positioned at 4.268 $\mu$m and 4.380 $\mu$m respectively. Finally, bands at 15.15 $\mu$m and 15.27 are attributed to O=C=O bending ($\nu_2$), while the 2.697 $\mu$m and 2.778 $\mu$m bands are attributed to $\nu_1$+$\nu_3$ and 2$\nu_2$+$\nu_3$ combinations respectively.

Two other ices of interest include hydrogen and nitrogen ice -- both of which have been suggested for the composition of `Oumuamua. Hydrogen ice has no prominent emission lines, and nitrogen ice has a weak absorption band at 2.15 $\mu$m, with the implication that these substances would be challenging to detect \citep{2020ApJ...896L...8S, 2013Icar..223..710G}. By contrast, nitrogen ice, as well as carbon monoxide and methane ices, are readily identifiable in reflectance spectra of Pluto and Triton taken by the SpeX spectrograph and imager aboard NASA's Infrared Telescope Facility \citep{2013Icar..223..710G, 2010Icar..205..594G}.

Finally, ammonia ice has absorption bands at 2.006 and 2.229 $\mu$m attributed to v$_3+$v$_4$ v$_3+$v$_2$ and combination modes respectively \citep{2009ApJS..181...53Z}. If ammonia ice is mixed with water ice, the wavelength of these absorption bands will shift towards shorter wavelengths; a higher percentage of water ice results in a larger shift towards shorter the wavelengths.

\begin{figure*}
\begin{center}
    \includegraphics[scale=0.4]{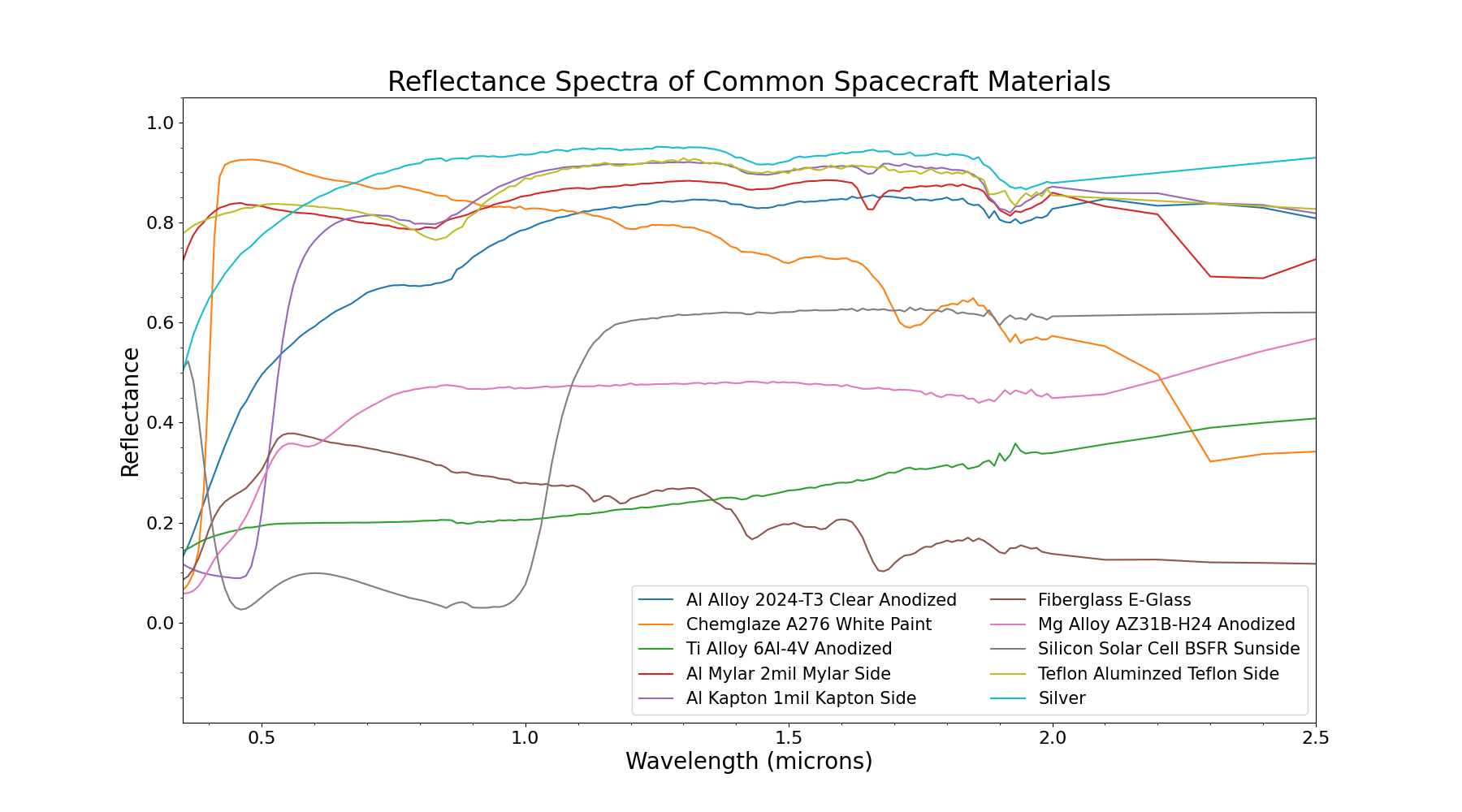}
\end{center}

\caption{Reflectance spectra of commonly-used spacecraft materials measured in the laboratory over the UV to shortwave-IR spectral range ($\sim$0.4 to 2.5 microns). Each artificial material shows a distinctive spectral signature that can be used for its detection and/or identification. (Source: Mark Elowitz)}
\label{fig:artificial_spectra}
\end{figure*}

\subsection{Artificial materials}

We may anticipate that the spectra of artificial materials of extrasolar provenance may exhibit marked differences with respect to both naturally occurring and human-manufactured materials.  Nevertheless, these latter materials provide the best analogue available and so we briefly review their significant features. Reflectance spectra of common spacecraft materials are visualized in Figure \ref{fig:artificial_spectra}.

\cite{2021JAnSc..68.1186C} created DebriSat, a high-fidelity, mock-up satellite representative of present-day, low Earth orbit satellites and is constructed with modern spacecraft materials of various shapes, bulk density, and dimensions. It is tested in controlled and instrumented laboratory conditions to see how the materials on board change with space conditions. The study showcases spectra measured from solar cells, circuit boards, and aluminum 6061 (Al 6061). Al 6061 is the most commonly used aluminum alloy and is typically used in aircraft components and hardware (e.g., cameras and electrical components and connectors).

Spectra were taken in NASA's Optical Measurement Center with an Analytical Spectral Device (ASD) field spectrometer with a range from 350 - 2500 nm. In the optical regime (350-700 nm), the blue anodized Al 6061 sample had a peak in reflectance near 400 nm. The gold and red Al 6061 had uneven peaks in reflectance near 550-600 nm. Magenta's purple and red components were represented in the spectra of the magenta Al 6061 sample with two reflectance peaks near 380 nm and 600 nm. The black Al 6061 sample was featureless and had low reflectance throughout the entire visible region. In the infrared (beyond 700 nm), all samples had an absorption feature at 850 nm. The absorption of the anodized/unpainted Al 6061 sample was the deepest, likely due to the lack of interference from the absence of paint. Additionally, there were absorption features at 1400 and 1900 nm that were likely the result of water. The last feature of note was the wide absorption at 2200 nm, indicating the presence of organic content. This was likely from C-H bonds or oxygen-hydrogen bonds.

While observing artificial materials in controlled environments is instructive (e.g., DebriSat), studies of materials exposed to uncontrolled, natural space environments can tell us how these materials respond to the space environment. \cite{2009ESASP.672E..42A} captured spectra at the NASA Infrared Telescope Facility of objects in GEO (geostationary orbit) or near-GEO which is ~36,000 km above the surface. Their results included a comparison of spectra of a spacecraft, a rocket body, and debris (see Figure 9 in \cite{2009ESASP.672E..42A}). The spacecraft is a Hughes 376 bus type - a cylindrical satellite with a de-spun dish, covered with solar cells around the body. The rocket body (R/B) under observation was the Inertial Upper Stage (IUS) with outer surfaces covered in white paint, multi-layer insulation, and carbon-carbon epoxy. The debris was from the Titan 3C Transtage breakup.

The spectra of the R/B and Titan debris increase in slope through 1.5 $\mu$m. The spacecraft spectrum has a notable band gap near 1 $\mu$m. Additionally, C-H absorption features at about 2.3 $\mu$m are present for both the spacecraft and R/B. For the spacecraft, this is thought to be from the C-H bonds in the solar cell material; on the other hand, for the R/B, the absorption is likely due to reflective paint. The centers of the 2.3 $\mu$m absorption are slightly shifted from one another. This shift is also seen in the spectra of laboratory samples of white paint and solar cells, and is therefore not an effect of the space environment. By looking at the slopes and absorption features of these three spectra, the study concluded that each class of object (spacecraft, R/B, debris) has a distinctive, readily distinguishable spectrum.

As in natural materials, multiple artificial materials may be present and disentangling their spectral signatures can be a challenge, partly because some features overlap.  Carbon monoxide's absorption feature at 2.35 $\mu$m is close to methane's in this part of the spectrum. The spacecraft and R/B in the study by \cite{2009ESASP.672E..42A} both had absorption features around 2.3 $\mu$m as well.

\subsection{Albedo}

\begin{figure}[h!]
\includegraphics[scale=0.71]{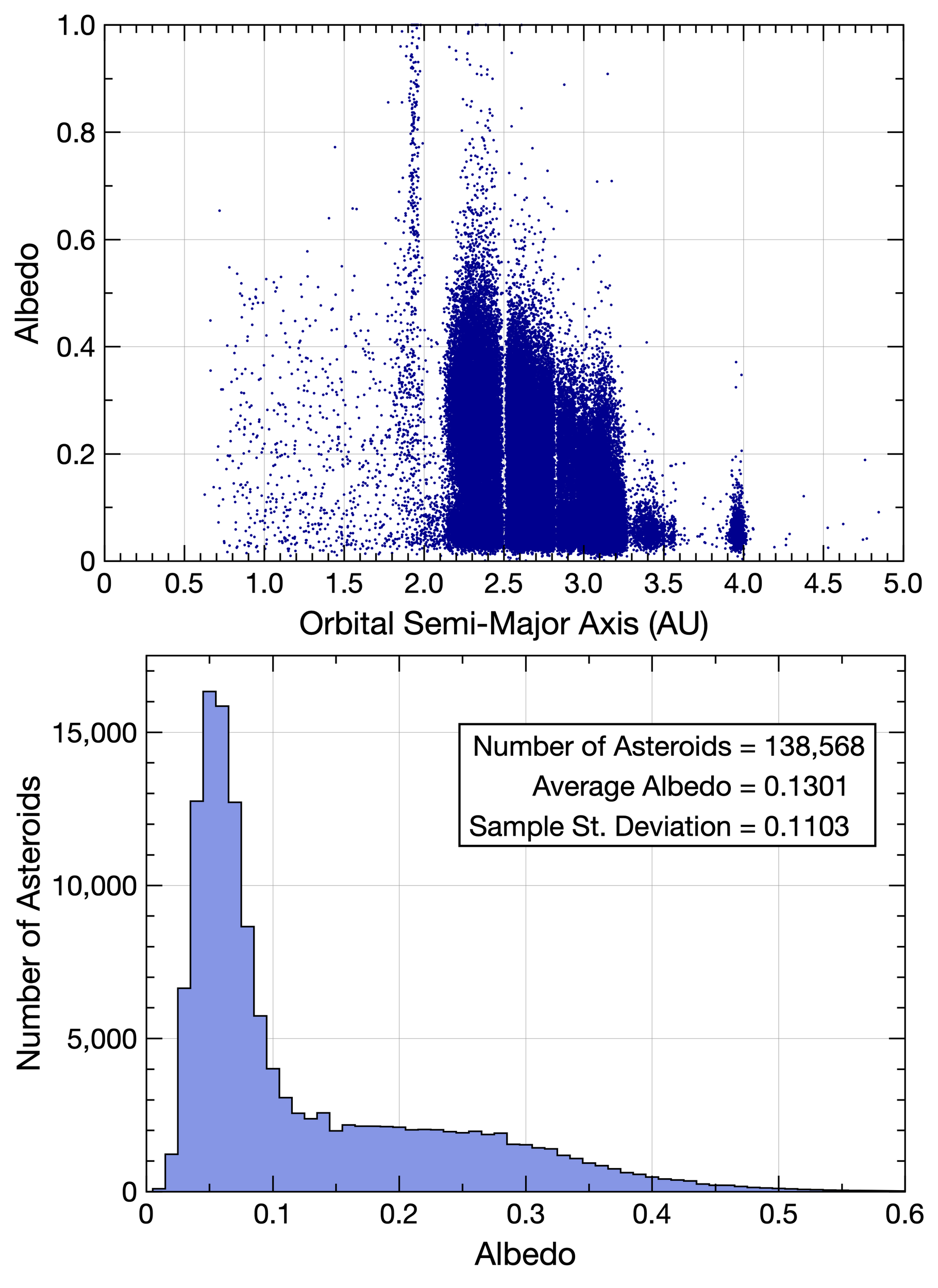}

\caption{{\em Top}:~Albedo of 138,568 Solar System asteroids versus their
  orbital semi-major axis.
  {\em Bottom}:~Histogram of asteroid albedos.  (Source:~JPL Small-Body
  Database, 2022, https://ssd.jpl.nasa.gov/tools/sbdb\_query.html)}

\label{fig:asteroid_albedo}
\vspace{-1.2mm}
\end{figure}

Another possible discriminator between a natural object, such as an asteroid or comet, and an artificial object of extraterrestrial origin is its Bond albedo -- the fraction of the total incident solar radiation reflected by an object back to space.  As shown in Fig.$\:$\ref{fig:asteroid_albedo}, the average albedo of Solar System asteroids is 0.13$\,\pm\,$0.11, while that of 19 Solar System comets is 0.05$\,\pm\,$0.03 (JPL Small-Body Database, 2022).  By contrast, the albedo of metallic surfaces, as might apply to spacecraft, typically range between 0.30 and 0.98 (e.g., \citep{mongelli2019albedo}).

\begin{figure}[b!]
\centering
\includegraphics[scale=0.5]{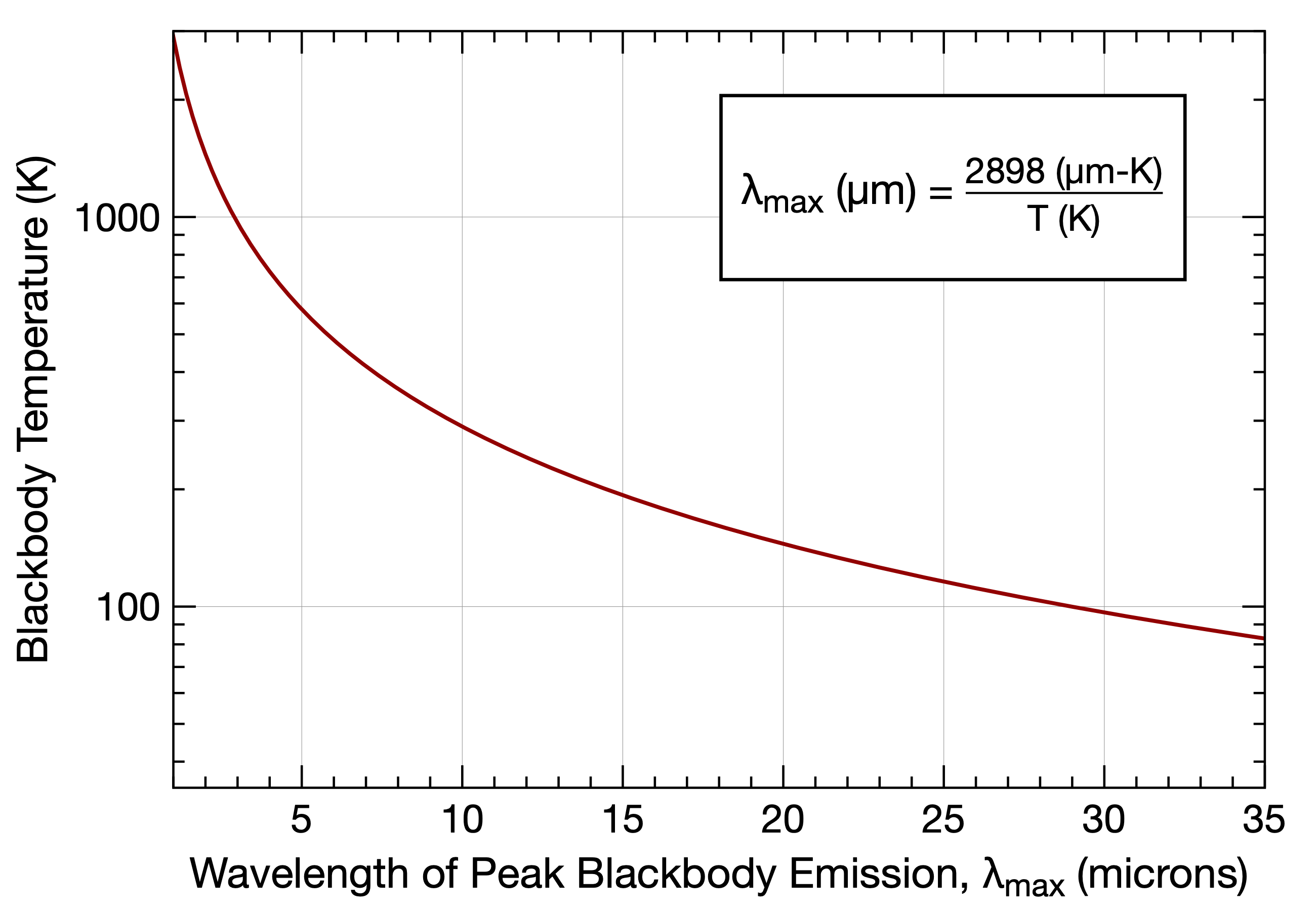}
\caption{The relation between an object's blackbody temperature and the
  wavelength at which its emission peaks (i.e., Wien's Law).}
\label{fig:wien}
\vspace{-1.2mm}
\end{figure}

For an ISO of undetermined size, which is likely the case when an object is first detected, the albedo can be derived with knowledge of its equilibrium temperature and distance from the Sun.  The ISO temperature can be determined from its spectrum -- specifically, the wavelength at which its continuum emission peaks, as shown in Fig.$\:$\ref{fig:wien} -- while its distance from the Sun must be inferred from a trajectory determination.

Assuming the ISO to be a spherical blackbody of uniform surface temperature, the Bond albedo, $a$, is given by


\begin{eqnarray}
a  & = & 4\,\times\,\left[ 0.25 - \left( \frac{T_{ISO}}{T_{\odot}}\,\sqrt{\frac{d}{R_{\odot}}}\: \right)^{\!\!\!4}\; \right] \nonumber \\*[3.0mm]
 & = & 4\,\times\,\left[ 0.25 - 4.09\,\times\,10^{-11} \left( T_{ISO}\,\sqrt{d_{AU}} \right)^{\!4}\: \right]\,,
\end{eqnarray}

\vspace{1.6mm}

\noindent where $T_{ISO}$ is the spectrum-determined ISO temperature in Kelvins, $T_{\odot}$ is the Sun's effective blackbody temperature of 5780$\:$K, $d$ is the ISO distance from the Sun in cm, $d_{AU}$ is the ISO distance from the Sun in AU, and $R_{\odot}$ is the Sun's radius of 7$\,\times\,$10$^{10}$ cm.

To account for departures from sphericity and a uniform surface temperature, it will be important to obtain spectra over multiple days.  The periodicity and range of albedo variations thus derived can provide important information about the object's rotation and surface composition while helping to differentiate between natural and artificial objects.

\subsection{Polarimetry}

The identification of the specific type of object (e.g. natural solar system object, artificial space probe) can be augmented when polarimetry measurements are used in synergy with spectral data. Polarimetric observations using an instrument capable of recording the intensity of light at several different orientations (from which the Stokes parameters can be estimated) can provide information about the surface properties (e.g. roughness, orientation) of an unknown object. For example, smooth/shinny metallic alloys tend to produce a single surface scatter leading to a linear polarization of the return signal to the observing sensor in contrast to regolith, which tends to produce a more complex volumetric scattering that does not produce strong polarization. The degree of polarization can be quantified using the derived Stokes parameters from the recorded intensity data by computing the degree of linear polarization. If the polarization data indicates a metallic surface, spectroscopy can be used to potentially identify the specific metal or metal alloy (e.g. the 0.84 micron absorption band indicative of aluminum).

\subsection{Impact features}

Impact cratering contributes to surface roughness, even on metallic surfaces exposed to space for long periods, which may limit the utility of polarimetric observations as a discriminating observation. The size distribution of impact features will vary dramatically depending on the provenance of ISOs in extrasolar systems.  Collision rates in peripheral Oort clouds are relatively low, as evidenced by the New Horizons close encounter with the KBO 486958 Arrokoth \citep{doi:10.1126/science.aaw9771}.  Even inner solar system objects may not preserve small-scale roughness from impacts on account of impact-induced seismic shaking, which tends to eliminate these features \citep{doi:10.1126/science.1104731}.  A solar sail ISO made of a thin sheet may be densely perforated by impacts and for this reason exhibit nonzero bulk transmissivity.  A camera prepositioned to capture the object following closest approach during a spacecraft rendezvous may permit occulation observations of background stars and determine whether starlight is transmitted through the object.

\subsection{Imaging spectrometry}

A future space mission designed to investigate the nature of interstellar objects (natural and/or artificial) should carry a visible/near-IR imaging spectrometer, a MWIR/LWIR imaging spectrometer, and an imaging polarimeter. The spectrometer(s) and polarimeter will be pointed in the direction of motion in order to collect a long-duration measurement, although we note that spectral imaging could present a challenge (for narrow-field observations in the flux-limited regime). If the main objective of the mission is to discriminate a possible artificial interstellar probe from a natural asteroid or cometary object, then an VNIR/SWIR imaging spectrometer sensitive to the wavelength range of 0.4 to 2.5 $\mu$m may be sufficient, based on reference spectra of various artificial and natural minerals. Advanced spectral mapping techniques, such as principal components analysis, minimum noise fraction transforms, matched filter based methods, linear mixing, and spectral feature fitting could be used to produce a geochemical survey map of the object. A seldom used, but extremely powerful technique known as derivative spectroscopy would accentuate any subtle, weak, or blended spectral features present in each pixel of the spectral cube data. False-color maps of the icy satellite surfaces based on derivative spectral data would tend to enhance different minerals and/or artificial materials present in the surface layers of the object. These proven methods have been extensively applied in the terrestrial remote sensing community \citep{kale2017research}. The data from the spectral and polarimetric instruments, when used in synergy, would provide the capability to assess if the material properties of the interstellar object differ markedly from natural solar system objects and/or human-made artificial materials.

\section{Mission Storage and Surveyor Telescope Locations}
\label{sec:missionstorage}

The optimum telescope  and survey strategy to select the very best ISO target for a one-shot interstellar object interceptor mission to visit requires an optimal storage location to be defined. L2 would be a reasonable place to discover and store the interstellar object interceptor mission since it is the site for major telescopes (JWST, now operating splendidly and the Roman Wide-Field telescope to be placed there by 2028). L2 is a stable point in space defined by the Earth - Sun system that allows nearly full sky viewing (excluding $<40$ degrees from Sun) and which has fully developed data download and mission commanding from the NASA Deep Space Network. L2 is also the preferred location for a Probe-class NASA mission to locate a 1.5m telescope for the Time-domain Spectroscopic Observatory (TSO) for  imaging and spectroscopy over the broad band 0.3 - 5  (microns). TSO would be the first space telescope with rapid response to study very high redshift gamma-ray bursts (GRBs) back to the very first generation of massive stars (PopIII) at redshifts z = 10 - 25. 

TSO (L2) is relevant to the interstellar object interceptor mission because a key mission requirement is to provide the most complete and compelling set of measurements prior to launch. The best astrometry and photometry for orbit determination and measurements of non-gravitational acceleration requires \textit{two} telescopes: TSO at L2 and the now-planned Near Earth Object (NEO)-Surveyor (formerly called NEOCam), at L1 are required to discover the ``best" ISO for the interstellar object interceptor mission to explore. The first advantage is L2 and L1, combined, provide essentially full sky coverage with minimal seasonal (Sun) loss of coverage. The second advantage is the L1 vs. L2 baseline provides the best astrometry for orbit determination and best parallax and distance measures. Only ISOs arriving on a trajectory that is roughly aligned with the L1-L2 axis will be missed, if only temporarily.

Unlike JWST, which takes 2 - 5 days to plan and then point to different targets, TSO would enable rapid (3 - 10 min) slews for rapid near-UV to mid-IR imaging and spectra of transients, particularly Gamma-ray Bursts (GRBs), but also all classes of transients, from flaring M-dwarfs to blazars to tumbling or rotating ISOs. TSO would have very deep imaging sensitivity: AB magnitude 25 in a 5 min exposure due to its very cold (100K) mirror, passively cooled by Sun-shields and radiating into black space.  No ground-based telescope (including ELTs) can deliver comparable performance. TSO at L2, together with the NEO Surveyor at L1, will be used to search for dangerous NEO's (including any ISOs) coming from the Sunward direction. This would allow optimum parallax and orbit determination for any/all ISOs and then maximal light curve coverage simultaneously in several bands and using moderate resolution spectroscopy. TSO's capabilites are described in \cite{2019BAAS...51c.548G}, \cite{2019BAAS...51c.472T}, \cite{2019astro2020T.306M}, and \cite{2019BAAS...51c.274S}.

Athough hardware at L2 must expend fuel to avoid drifting away from this unstable position, a spacecraft can sit at GEO without expending much, if any, fuel.  Since intercepting an ISO may require significant amounts of propulsion, preserving fuel before committing to an intercept could be critical.  The trade to be considered is the fuel required to climb out of the earth's gravity well from GEO (or, maybe even lunar orbit) versus the fuel required to station keep at L2 for a year (or more). The L2 site is preferred (despite station-keeping costs) since it has significantly larger sky coverage than GEO (with Earth, Moon and Sun restrictions) and provides longer-term storage possibilities. 

\section{Conclusions}
\label{sec:conclusion}

We have explored physical considerations involved in planning an intercept mission to an `Oumuamua-like object. We found that a close encounter with an ISO at a distance of $\lesssim 10^3 \mathrm{\; km}$ would allow for  high-resolution measurements, but that requiring the impact parameter to be $\gtrsim 10^2 \mathrm{\; km}$ would ensure sufficient time for taking observations. We predicted that LSST will discover $\sim 15$ `Oumuamua-like objects over its 10-year lifespan, with a 95\% Poisson confidence interval spanning the range 0.38 – 84. We calculated that a mission with a budget of $\Delta v_{UL} \sim 30 \mathrm{\; km \; s^{-1}}$ would have an $\sim 85 \%$ chance of suitable `Oumuamua-like object in 10-years, but a $\gtrsim 90 \%$ chance of finding an ISO a third smaller in size, although such a small object would require launch to follow almost immediately after detection. We then identified surface properties and spectral features of rocky materials, ices, and arificial (manmade) materials, as well as expectations for albedo and polarimetry measurements, that can be used to discriminate between categories of interstellar objects. Finally, we discussed imaging spectrometry capabilities and storage locations for an interstellar object mission.

\clearpage




\bibliography{sample63}{}

\begin{thebibliography}{}
\expandafter\ifx\csname natexlab\endcsname\relax\def\natexlab#1{#1}\fi
\providecommand{\url}[1]{\href{#1}{#1}}
\providecommand{\dodoi}[1]{doi:~\href{http://doi.org/#1}{\nolinkurl{#1}}}
\providecommand{\doeprint}[1]{\href{http://ascl.net/#1}{\nolinkurl{http://ascl.net/#1}}}
\providecommand{\doarXiv}[1]{\href{https://arxiv.org/abs/#1}{\nolinkurl{https://arxiv.org/abs/#1}}}

\bibitem[{{Abercromby} {et~al.}(2009){Abercromby}, {Abell}, \&
  {Barker}}]{2009ESASP.672E..42A}
{Abercromby}, K.~J., {Abell}, P., \& {Barker}, E. 2009, in ESA Special
  Publication, Vol. 672, Fifth European Conference on Space Debris, ed.
  H.~{Lacoste}, 42

\bibitem[{{Almeida-Fernandes} \& {Rocha-Pinto}(2018)}]{Fernandes2018}
{Almeida-Fernandes}, F., \& {Rocha-Pinto}, H.~J. 2018, \mnras, 480, 4903,
  \dodoi{10.1093/mnras/sty2202}

\bibitem[{{Aravind} {et~al.}(2021){Aravind}, {Ganesh}, {Venkataramani}, {Sahu},
  {Angchuk}, {Sivarani}, \& {Unni}}]{Aravind2021}
{Aravind}, K., {Ganesh}, S., {Venkataramani}, K., {et~al.} 2021, \mnras, 502,
  3491, \dodoi{10.1093/mnras/stab084}

\bibitem[{{Bagnulo} {et~al.}(2021){Bagnulo}, {Cellino}, {Kolokolova},
  {Ne{\v{z}}i{\v{c}}}, {Santana-Ros}, {Borisov}, {Christou}, {Bendjoya}, \&
  {Devog{\`e}le}}]{Bagnulo2021}
{Bagnulo}, S., {Cellino}, A., {Kolokolova}, L., {et~al.} 2021, Nature
  Communications, 12, 1797, \dodoi{10.1038/s41467-021-22000-x}

\bibitem[{{Bannister} {et~al.}(2017){Bannister}, {Schwamb}, {Fraser},
  {Marsset}, {Fitzsimmons}, {Benecchi}, {Lacerda}, {Pike}, {Kavelaars},
  {Smith}, {Stewart}, {Wang}, \& {Lehner}}]{Bannister2017}
{Bannister}, M.~T., {Schwamb}, M.~E., {Fraser}, W.~C., {et~al.} 2017, \apjl,
  851, L38, \dodoi{10.3847/2041-8213/aaa07c}

\bibitem[{{Bannister} {et~al.}(2020){Bannister}, {Opitom}, {Fitzsimmons},
  {Moulane}, {Jehin}, {Seligman}, {Rousselot}, {Knight}, {Marsset}, {Schwamb},
  {Guilbert-Lepoutre}, {Jorda}, {Vernazza}, \& {Benkhaldoun}}]{Bannister2020}
{Bannister}, M.~T., {Opitom}, C., {Fitzsimmons}, A., {et~al.} 2020, arXiv
  e-prints, arXiv:2001.11605.
\newblock \doarXiv{2001.11605}

\bibitem[{{Belton} {et~al.}(2018){Belton}, {Hainaut}, {Meech}, {Mueller},
  {Kleyna}, {Weaver}, {Buie}, {Drahus}, {Guzik}, {Wainscoat}, {Waniak},
  {Handzlik}, {Kurowski}, {Xu}, {Sheppard}, {Micheli}, {Ebeling}, \&
  {Keane}}]{Belton2018}
{Belton}, M.~J.~S., {Hainaut}, O.~R., {Meech}, K.~J., {et~al.} 2018, ApJL, 856,
  L21, \dodoi{10.3847/2041-8213/aab370}

\bibitem[{Bialy \& Loeb(2018)}]{bialy2018radiation}
Bialy, S., \& Loeb, A. 2018, \apjl, 868, L1

\bibitem[{{Bodewits} {et~al.}(2020){Bodewits}, {Noonan}, {Feldman},
  {Bannister}, {Farnocchia}, {Harris}, {Li}, {Mandt}, {Parker}, \&
  {Xing}}]{Bodewits2020}
{Bodewits}, D., {Noonan}, J.~W., {Feldman}, P.~D., {et~al.} 2020, Nature
  Astronomy, 4, 867, \dodoi{10.1038/s41550-020-1095-2}

\bibitem[{{Bolin} {et~al.}(2020{\natexlab{a}}){Bolin}, {Bodewits}, {Lisse},
  {Fernandez}, {Helou}, \& {Cenko}}]{Bolin2020ATel}
{Bolin}, B.~T., {Bodewits}, D., {Lisse}, C.~M., {et~al.} 2020{\natexlab{a}},
  The Astronomer's Telegram, 13613, 1

\bibitem[{{Bolin} {et~al.}(2018{\natexlab{a}}){Bolin}, {Weaver}, {Fernandez},
  {Lisse}, {Huppenkothen}, {Jones}, {Juri{\'c}}, {Moeyens}, {Schambeau},
  {Slater}, {Ivezi{\'c}}, \& {Connolly}}]{Bolin2017}
{Bolin}, B.~T., {Weaver}, H.~A., {Fernandez}, Y.~R., {et~al.}
  2018{\natexlab{a}}, ApJL, 852, L2, \dodoi{10.3847/2041-8213/aaa0c9}

\bibitem[{{Bolin} {et~al.}(2018{\natexlab{b}}){Bolin}, {Weaver}, {Fernandez},
  {Lisse}, {Huppenkothen}, {Jones}, {Juri{\'c}}, {Moeyens}, {Schambeau},
  {Slater}, {Ivezi{\'c}}, \& {Connolly}}]{Bolin2018}
---. 2018{\natexlab{b}}, \apjl, 852, L2, \dodoi{10.3847/2041-8213/aaa0c9}

\bibitem[{{Bolin} {et~al.}(2020{\natexlab{b}}){Bolin}, {Lisse}, {Kasliwal},
  {Quimby}, {Tan}, {Copperwheat}, {Lin}, {Morbidelli}, {Abe}, {Bendjoya},
  {Burdge}, {Coughlin}, {Fremling}, {Itoh}, {Koss}, {Masci}, {Maeno},
  {Mamajek}, {Marocco}, {Murata}, {Rivet}, {Sitko}, {Stern}, {Vernet},
  {Walters}, {Yan}, {Andreoni}, {Bhalerao}, {Bodewits}, {De}, {Deshmukh},
  {Bellm}, {Blagorodnova}, {Buzasi}, {Cenko}, {Chang}, {Chojnowski}, {Dekany},
  {Duev}, {Graham}, {Juri{\'c}}, {Kulkarni}, {Kupfer}, {Mahabal}, {Neill},
  {Ngeow}, {Penprase}, {Riddle}, {Rodriguez}, {Smith}, {Rosnet}, {Sollerman},
  \& {Soumagnac}}]{Bolin2019}
{Bolin}, B.~T., {Lisse}, C.~M., {Kasliwal}, M.~M., {et~al.} 2020{\natexlab{b}},
  \aj, 160, 26, \dodoi{10.3847/1538-3881/ab9305}

\bibitem[{{Boogert} {et~al.}(1997){Boogert}, {Schutte}, {Helmich}, {Tielens},
  \& {Wooden}}]{1997A&A...317..929B}
{Boogert}, A.~C.~A., {Schutte}, W.~A., {Helmich}, F.~P., {Tielens},
  A.~G.~G.~M., \& {Wooden}, D.~H. 1997, \aap, 317, 929

\bibitem[{Brunetto {et~al.}(2006)Brunetto, Vernazza, Marchi, Birlan,
  Fulchignoni, Orofino, \& Strazzulla}]{brunetto2006modeling}
Brunetto, R., Vernazza, P., Marchi, S., {et~al.} 2006, Icarus, 184, 327

\bibitem[{Castillo-Rogez {et~al.}(2019)Castillo-Rogez, Landau, Chung, \&
  Meech}]{Castillo-Rogez2019}
Castillo-Rogez, J., Landau, D., Chung, S.-J., \& Meech, K. 2019, in Approach to
  exploring interstellar objects and long-period comets

\bibitem[{{Cook} {et~al.}(2016){Cook}, {Ragozzine}, {Granvik}, \&
  {Stephens}}]{Cook2016}
{Cook}, N.~V., {Ragozzine}, D., {Granvik}, M., \& {Stephens}, D.~C. 2016, ApJ,
  825, 51, \dodoi{10.3847/0004-637X/825/1/51}

\bibitem[{{Cordiner} {et~al.}(2020){Cordiner}, {Milam}, {Biver},
  {Bockel{\'e}e-Morvan}, {Roth}, {Bergin}, {Jehin}, {Remijan}, {Charnley},
  {Mumma}, {Boissier}, {Crovisier}, {Paganini}, {Kuan}, \&
  {Lis}}]{Cordiner2020}
{Cordiner}, M.~A., {Milam}, S.~N., {Biver}, N., {et~al.} 2020, Nature
  Astronomy, 4, 861, \dodoi{10.1038/s41550-020-1087-2}

\bibitem[{{Cowardin} {et~al.}(2021){Cowardin}, {Hostetler}, {Murray}, {Reyes},
  \& {Cruz}}]{2021JAnSc..68.1186C}
{Cowardin}, H.~M., {Hostetler}, J.~M., {Murray}, J.~I., {Reyes}, J.~A., \&
  {Cruz}, C.~L. 2021, Journal of the Astronautical Sciences, 68, 1186,
  \dodoi{10.1007/s40295-021-00278-9}

\bibitem[{{Cremonese} {et~al.}(2020){Cremonese}, {Fulle}, {Cambianica},
  {Munaretto}, {Capria}, {La Forgia}, {Lazzarin}, {Migliorini}, {Boschin},
  {Milani}, {Aletti}, {Arlic}, {Bacci}, {Bacci}, {Bryssinck}, {Carosati},
  {Castellano}, {Buzzi}, {Di Rubbo}, {Facchini}, {Guido}, {Kugel}, {Ligustri},
  {Maestripieri}, {Mantero}, {Nicolas}, {Ochner}, {Perrella}, {Trabatti}, \&
  {Valvasori}}]{Cremonese2020}
{Cremonese}, G., {Fulle}, M., {Cambianica}, P., {et~al.} 2020, \apjl, 893, L12,
  \dodoi{10.3847/2041-8213/ab8455}

\bibitem[{{de la Fuente Marcos} \& {de la Fuente
  Marcos}(2020)}]{delafuente2020}
{de la Fuente Marcos}, C., \& {de la Fuente Marcos}, R. 2020, \aap, 643, A18,
  \dodoi{10.1051/0004-6361/202037447}

\bibitem[{Desch \& Jackson(2021)}]{desch20211i}
Desch, S.~J., \& Jackson, A.~P. 2021, Journal of Geophysical Research: Planets,
  e2020JE006807

\bibitem[{{Do} {et~al.}(2018){Do}, {Tucker}, \& {Tonry}}]{Do2018}
{Do}, A., {Tucker}, M.~A., \& {Tonry}, J. 2018, \apjl, 855, L10,
  \dodoi{10.3847/2041-8213/aaae67}

\bibitem[{Donitz {et~al.}(2021)Donitz, Castillo-Rogez, \&
  Matousek}]{Donitz2021}
Donitz, B.~P., Castillo-Rogez, J.~C., \& Matousek, S.~E. 2021, in 2021 IEEE
  Aerospace Conference (50100), 1--12, \dodoi{10.1109/AERO50100.2021.9438142}

\bibitem[{{Drahus} {et~al.}(2018){Drahus}, {Guzik}, {Waniak}, {Handzlik},
  {Kurowski}, \& {Xu}}]{Drahus2018}
{Drahus}, M., {Guzik}, P., {Waniak}, W., {et~al.} 2018, Nature Astronomy, 2,
  407, \dodoi{10.1038/s41550-018-0440-1}

\bibitem[{{Drahus} {et~al.}(2020){Drahus}, {Guzik}, {Udalski}, {Ratajczak},
  {Rybicki}, {Iwanek}, {Wrona}, {Pietrukowicz}, {Soszynski}, {Ulaczyk},
  {Poleski}, {Mroz}, {Szymanski}, {Skowron}, {Gromadzki}, {Skowron}, \&
  {Kozlowski}}]{Drahus2020ATel}
{Drahus}, M., {Guzik}, P., {Udalski}, A., {et~al.} 2020, The Astronomer's
  Telegram, 13549, 1

\bibitem[{{Engelhardt} {et~al.}(2017){Engelhardt}, {Jedicke}, {Vere{\v{s}}},
  {Fitzsimmons}, {Denneau}, {Beshore}, \& {Meinke}}]{Engelhardt2014}
{Engelhardt}, T., {Jedicke}, R., {Vere{\v{s}}}, P., {et~al.} 2017, AJ, 153,
  133, \dodoi{10.3847/1538-3881/aa5c8a}

\bibitem[{{Feng} \& {Jones}(2018)}]{Feng2018}
{Feng}, F., \& {Jones}, H.~R.~A. 2018, \apjl, 852, L27,
  \dodoi{10.3847/2041-8213/aaa404}

\bibitem[{{Fitzsimmons} {et~al.}(2018){Fitzsimmons}, {Snodgrass}, {Rozitis},
  {Yang}, {Hyland}, {Seccull}, {Bannister}, {Fraser}, {Jedicke}, \&
  {Lacerda}}]{Fitzsimmons2017}
{Fitzsimmons}, A., {Snodgrass}, C., {Rozitis}, B., {et~al.} 2018, Nature
  Astronomy, 2, 133, \dodoi{10.1038/s41550-017-0361-4}

\bibitem[{{Fitzsimmons} {et~al.}(2019){Fitzsimmons}, {Hainaut}, {Meech},
  {Jehin}, {Moulane}, {Opitom}, {Yang}, {Keane}, {Kleyna}, {Micheli}, \&
  {Snodgrass}}]{Fitzsimmons:2019}
{Fitzsimmons}, A., {Hainaut}, O., {Meech}, K.~J., {et~al.} 2019, \apjl, 885,
  L9, \dodoi{10.3847/2041-8213/ab49fc}

\bibitem[{{Fraser} {et~al.}(2018){Fraser}, {Pravec}, {Fitzsimmons}, {Lacerda},
  {Bannister}, {Snodgrass}, \& {Smoli{\'c}}}]{Fraser2017}
{Fraser}, W.~C., {Pravec}, P., {Fitzsimmons}, A., {et~al.} 2018, Nature
  Astronomy, 2, 383, \dodoi{10.1038/s41550-018-0398-z}

\bibitem[{F{\"u}glistaler \& Pfenniger(2015)}]{fuglistaler2015solid}
F{\"u}glistaler, A., \& Pfenniger, D. 2015, A\&A, 578, A18

\bibitem[{{Gaidos} {et~al.}(2017){Gaidos}, {Williams}, \&
  {Kraus}}]{Gaidos2017a}
{Gaidos}, E., {Williams}, J., \& {Kraus}, A. 2017, RNAAS, 1, 13,
  \dodoi{10.3847/2515-5172/aa9851}

\bibitem[{{Gerakines} {et~al.}(1995){Gerakines}, {Schutte}, {Greenberg}, \&
  {van Dishoeck}}]{1995A&A...296..810G}
{Gerakines}, P.~A., {Schutte}, W.~A., {Greenberg}, J.~M., \& {van Dishoeck},
  E.~F. 1995, \aap, 296, 810.
\newblock \doarXiv{astro-ph/9409076}

\bibitem[{{Grindlay} {et~al.}(2019){Grindlay}, {Berger}, {Metzger}, {Gezari},
  {Ivezic}, {Jencson}, {Kasliwal}, {Kutyrev}, {Macleod}, {Melnick}, {Purcell},
  {Rieke}, {Shen}, {Tanvir}, \& {Vasey}}]{2019BAAS...51c.548G}
{Grindlay}, J., {Berger}, E., {Metzger}, B., {et~al.} 2019, \baas, 51, 548.
\newblock \doarXiv{1903.07828}

\bibitem[{{Grundy} {et~al.}(2013){Grundy}, {Olkin}, {Young}, {Buie}, \&
  {Young}}]{2013Icar..223..710G}
{Grundy}, W.~M., {Olkin}, C.~B., {Young}, L.~A., {Buie}, M.~W., \& {Young},
  E.~F. 2013, \icarus, 223, 710, \dodoi{10.1016/j.icarus.2013.01.019}

\bibitem[{{Grundy} {et~al.}(2010){Grundy}, {Young}, {Stansberry}, {Buie},
  {Olkin}, \& {Young}}]{2010Icar..205..594G}
{Grundy}, W.~M., {Young}, L.~A., {Stansberry}, J.~A., {et~al.} 2010, \icarus,
  205, 594, \dodoi{10.1016/j.icarus.2009.08.005}

\bibitem[{{Guzik} \& {Drahus}(2021)}]{Guzik2021}
{Guzik}, P., \& {Drahus}, M. 2021, \nat, 593, 375,
  \dodoi{10.1038/s41586-021-03485-4}

\bibitem[{{Guzik} {et~al.}(2020){Guzik}, {Drahus}, {Rusek}, {Waniak},
  {Cannizzaro}, \& {Pastor-Marazuela}}]{Guzik:2020}
{Guzik}, P., {Drahus}, M., {Rusek}, K., {et~al.} 2020, Nature Astronomy, 4, 53,
  \dodoi{10.1038/s41550-019-0931-8}

\bibitem[{Hallatt \& Wiegert(2020)}]{hallatt2020dynamics}
Hallatt, T., \& Wiegert, P. 2020, \aj, 159, 147

\bibitem[{Hapke(2001)}]{hapke2001space}
Hapke, B. 2001, Journal of Geophysical Research: Planets, 106, 10039

\bibitem[{{Hein} {et~al.}(2017){Hein}, {Perakis}, {Eubanks}, {Hibberd},
  {Crowl}, {Hayward}, {Kennedy}, \& {Osborne}}]{Hein2017}
{Hein}, A.~M., {Perakis}, N., {Eubanks}, T.~M., {et~al.} 2017, arXiv e-prints,
  arXiv:1711.03155.
\newblock \doarXiv{1711.03155}

\bibitem[{{Hibberd} {et~al.}(2022){Hibberd}, {Hein}, {Eubanks}, \&
  {Kennedy}}]{Hibberd2022}
{Hibberd}, A., {Hein}, A., {Eubanks}, M., \& {Kennedy}, Robert, I. 2022, arXiv
  e-prints, arXiv:2201.04240.
\newblock \doarXiv{2201.04240}

\bibitem[{{Hibberd} {et~al.}(2020){Hibberd}, {Hein}, \&
  {Eubanks}}]{Hibberd2020}
{Hibberd}, A., {Hein}, A.~M., \& {Eubanks}, T.~M. 2020, Acta Astronautica, 170,
  136, \dodoi{10.1016/j.actaastro.2020.01.018}

\bibitem[{Hoang \& Loeb(2020)}]{Hoang2020}
Hoang, T., \& Loeb, A. 2020, \apjl, 899, L23

\bibitem[{{Hoover} {et~al.}(2022){Hoover}, {Seligman}, \& {Payne}}]{Hoover2022}
{Hoover}, D.~J., {Seligman}, D.~Z., \& {Payne}, M.~J. 2022, \psj, 3, 71,
  \dodoi{10.3847/PSJ/ac58fe}

\bibitem[{{Hsieh} {et~al.}(2021){Hsieh}, {Laughlin}, \& {Arce}}]{Hsieh2021}
{Hsieh}, C.-H., {Laughlin}, G., \& {Arce}, H.~G. 2021, \apj, 917, 20,
  \dodoi{10.3847/1538-4357/ac0729}

\bibitem[{{Hui} {et~al.}(2020){Hui}, {Ye}, {F{\"o}hring}, {Hung}, \&
  {Tholen}}]{Hui2020}
{Hui}, M.-T., {Ye}, Q.-Z., {F{\"o}hring}, D., {Hung}, D., \& {Tholen}, D.~J.
  2020, \aj, 160, 92, \dodoi{10.3847/1538-3881/ab9df8}

\bibitem[{{Ivezi{\'c}} {et~al.}(2019){Ivezi{\'c}}, {Kahn}, {Tyson}, {Abel},
  {Acosta}, {Allsman}, {Alonso}, {AlSayyad}, {Anderson}, {Andrew}, {Angel},
  {Angeli}, {Ansari}, {Antilogus}, {Araujo}, {Armstrong}, {Arndt}, {Astier},
  {Aubourg}, {Auza}, {Axelrod}, {Bard}, {Barr}, {Barrau}, {Bartlett}, {Bauer},
  {Bauman}, {Baumont}, {Bechtol}, {Bechtol}, {Becker}, {Becla}, {Beldica},
  {Bellavia}, {Bianco}, {Biswas}, {Blanc}, {Blazek}, {Blandford}, {Bloom},
  {Bogart}, {Bond}, {Booth}, {Borgland}, {Borne}, {Bosch}, {Boutigny},
  {Brackett}, {Bradshaw}, {Brandt}, {Brown}, {Bullock}, {Burchat}, {Burke},
  {Cagnoli}, {Calabrese}, {Callahan}, {Callen}, {Carlin}, {Carlson},
  {Chandrasekharan}, {Charles-Emerson}, {Chesley}, {Cheu}, {Chiang}, {Chiang},
  {Chirino}, {Chow}, {Ciardi}, {Claver}, {Cohen-Tanugi}, {Cockrum}, {Coles},
  {Connolly}, {Cook}, {Cooray}, {Covey}, {Cribbs}, {Cui}, {Cutri}, {Daly},
  {Daniel}, {Daruich}, {Daubard}, {Daues}, {Dawson}, {Delgado}, {Dellapenna},
  {de Peyster}, {de Val-Borro}, {Digel}, {Doherty}, {Dubois},
  {Dubois-Felsmann}, {Durech}, {Economou}, {Eifler}, {Eracleous}, {Emmons},
  {Fausti Neto}, {Ferguson}, {Figueroa}, {Fisher-Levine}, {Focke}, {Foss},
  {Frank}, {Freemon}, {Gangler}, {Gawiser}, {Geary}, {Gee}, {Geha}, {Gessner},
  {Gibson}, {Gilmore}, {Glanzman}, {Glick}, {Goldina}, {Goldstein}, {Goodenow},
  {Graham}, {Gressler}, {Gris}, {Guy}, {Guyonnet}, {Haller}, {Harris},
  {Hascall}, {Haupt}, {Hernandez}, {Herrmann}, {Hileman}, {Hoblitt}, {Hodgson},
  {Hogan}, {Howard}, {Huang}, {Huffer}, {Ingraham}, {Innes}, {Jacoby}, {Jain},
  {Jammes}, {Jee}, {Jenness}, {Jernigan}, {Jevremovi{\'c}}, {Johns}, {Johnson},
  {Johnson}, {Jones}, {Juramy-Gilles}, {Juri{\'c}}, {Kalirai}, {Kallivayalil},
  {Kalmbach}, {Kantor}, {Karst}, {Kasliwal}, {Kelly}, {Kessler}, {Kinnison},
  {Kirkby}, {Knox}, {Kotov}, {Krabbendam}, {Krughoff}, {Kub{\'a}nek},
  {Kuczewski}, {Kulkarni}, {Ku}, {Kurita}, {Lage}, {Lambert}, {Lange},
  {Langton}, {Le Guillou}, {Levine}, {Liang}, {Lim}, {Lintott}, {Long},
  {Lopez}, {Lotz}, {Lupton}, {Lust}, {MacArthur}, {Mahabal}, {Mandelbaum},
  {Markiewicz}, {Marsh}, {Marshall}, {Marshall}, {May}, {McKercher}, {McQueen},
  {Meyers}, {Migliore}, {Miller}, {Mills}, {Miraval}, {Moeyens}, {Moolekamp},
  {Monet}, {Moniez}, {Monkewitz}, {Montgomery}, {Morrison}, {Mueller},
  {Muller}, {Mu{\~n}oz Arancibia}, {Neill}, {Newbry}, {Nief}, {Nomerotski},
  {Nordby}, {O'Connor}, {Oliver}, {Olivier}, {Olsen}, {O'Mullane}, {Ortiz},
  {Osier}, {Owen}, {Pain}, {Palecek}, {Parejko}, {Parsons}, {Pease},
  {Peterson}, {Peterson}, {Petravick}, {Libby Petrick}, {Petry},
  {Pierfederici}, {Pietrowicz}, {Pike}, {Pinto}, {Plante}, {Plate}, {Plutchak},
  {Price}, {Prouza}, {Radeka}, {Rajagopal}, {Rasmussen}, {Regnault}, {Reil},
  {Reiss}, {Reuter}, {Ridgway}, {Riot}, {Ritz}, {Robinson}, {Roby}, {Roodman},
  {Rosing}, {Roucelle}, {Rumore}, {Russo}, {Saha}, {Sassolas}, {Schalk},
  {Schellart}, {Schindler}, {Schmidt}, {Schneider}, {Schneider}, {Schoening},
  {Schumacher}, {Schwamb}, {Sebag}, {Selvy}, {Sembroski}, {Seppala}, {Serio},
  {Serrano}, {Shaw}, {Shipsey}, {Sick}, {Silvestri}, {Slater}, {Smith},
  {Smith}, {Sobhani}, {Soldahl}, {Storrie-Lombardi}, {Stover}, {Strauss},
  {Street}, {Stubbs}, {Sullivan}, {Sweeney}, {Swinbank}, {Szalay}, {Takacs},
  {Tether}, {Thaler}, {Thayer}, {Thomas}, {Thornton}, {Thukral}, {Tice},
  {Trilling}, {Turri}, {Van Berg}, {Vanden Berk}, {Vetter}, {Virieux},
  {Vucina}, {Wahl}, {Walkowicz}, {Walsh}, {Walter}, {Wang}, {Wang}, {Warner},
  {Wiecha}, {Willman}, {Winters}, {Wittman}, {Wolff}, {Wood-Vasey}, {Wu},
  {Xin}, {Yoachim}, \& {Zhan}}]{Ivezic2019}
{Ivezi{\'c}}, {\v{Z}}., {Kahn}, S.~M., {Tyson}, J.~A., {et~al.} 2019, \apj,
  873, 111, \dodoi{10.3847/1538-4357/ab042c}

\bibitem[{Jackson \& Desch(2021)}]{jackson20211i}
Jackson, A.~P., \& Desch, S.~J. 2021, Journal of Geophysical Research: Planets,
  e2020JE006706

\bibitem[{{Jewitt} {et~al.}(2020{\natexlab{a}}){Jewitt}, {Hui}, {Kim},
  {Mutchler}, {Weaver}, \& {Agarwal}}]{Jewitt2020}
{Jewitt}, D., {Hui}, M.-T., {Kim}, Y., {et~al.} 2020{\natexlab{a}}, \apjl, 888,
  L23, \dodoi{10.3847/2041-8213/ab621b}

\bibitem[{{Jewitt} {et~al.}(2020{\natexlab{b}}){Jewitt}, {Kim}, {Mutchler},
  {Weaver}, {Agarwal}, \& {Hui}}]{Jewitt2020:BorisovBreakup}
{Jewitt}, D., {Kim}, Y., {Mutchler}, M., {et~al.} 2020{\natexlab{b}}, \apjl,
  896, L39, \dodoi{10.3847/2041-8213/ab99cb}

\bibitem[{{Jewitt} \& {Luu}(2019)}]{Jewitt2019b}
{Jewitt}, D., \& {Luu}, J. 2019, \apjl, 886, L29,
  \dodoi{10.3847/2041-8213/ab530b}

\bibitem[{{Jewitt} {et~al.}(2017){Jewitt}, {Luu}, {Rajagopal}, {Kotulla},
  {Ridgway}, {Liu}, \& {Augusteijn}}]{Jewitt2017}
{Jewitt}, D., {Luu}, J., {Rajagopal}, J., {et~al.} 2017, ApJL, 850, L36,
  \dodoi{10.3847/2041-8213/aa9b2f}

\bibitem[{{Jewitt} {et~al.}(2020{\natexlab{c}}){Jewitt}, {Mutchler}, {Kim},
  {Weaver}, \& {Hui}}]{Jewitt2020ATel}
{Jewitt}, D., {Mutchler}, M., {Kim}, Y., {Weaver}, H., \& {Hui}, M.-T.
  2020{\natexlab{c}}, The Astronomer's Telegram, 13611, 1

\bibitem[{{Jones} \& {ESA Comet Interceptor Team}(2019)}]{jones2019}
{Jones}, G., \& {ESA Comet Interceptor Team}. 2019, {Comet Interceptor A
  Mission to a Dynamically New Solar System Object}.
\newblock
  \url{http://www.cometinterceptor.space/uploads/1/2/3/7/123778284/comet_interceptor_executive_summary.pdf}

\bibitem[{{Jones} {et~al.}(2009){Jones}, {Chesley}, {Connolly}, {Harris},
  {Ivezic}, {Knezevic}, {Kubica}, {Milani}, \& {Trilling}}]{jones2009lsst}
{Jones}, R.~L., {Chesley}, S.~R., {Connolly}, A.~J., {et~al.} 2009, Earth Moon
  and Planets, 105, 101, \dodoi{10.1007/s11038-009-9305-z}

\bibitem[{{Jones} {et~al.}(2018){Jones}, {Slater}, {Moeyens}, {Allen},
  {Axelrod}, {Cook}, {Ivezi{\'c}}, {Juri{\'c}}, {Myers}, \&
  {Petry}}]{Jones2018}
{Jones}, R.~L., {Slater}, C.~T., {Moeyens}, J., {et~al.} 2018, \icarus, 303,
  181, \dodoi{10.1016/j.icarus.2017.11.033}

\bibitem[{Kale {et~al.}(2017)Kale, Solankar, Nalawade, Dhumal, \&
  Gite}]{kale2017research}
Kale, K.~V., Solankar, M.~M., Nalawade, D.~B., Dhumal, R.~K., \& Gite, H.~R.
  2017, Proceedings of the national academy of sciences, India section a:
  physical sciences, 87, 541

\bibitem[{{Kareta} {et~al.}(2020){Kareta}, {Andrews}, {Noonan}, {Harris},
  {Smith}, {O'Brien}, {Sharkey}, {Reddy}, {Springmann}, {Lejoly}, {Volk},
  {Conrad}, \& {Veillet}}]{Kareta:2019}
{Kareta}, T., {Andrews}, J., {Noonan}, J.~W., {et~al.} 2020, \apjl, 889, L38,
  \dodoi{10.3847/2041-8213/ab6a08}

\bibitem[{{Kim} {et~al.}(2020){Kim}, {Jewitt}, {Mutchler}, {Agarwal}, {Hui}, \&
  {Weaver}}]{Kim2020}
{Kim}, Y., {Jewitt}, D., {Mutchler}, M., {et~al.} 2020, \apjl, 895, L34,
  \dodoi{10.3847/2041-8213/ab9228}

\bibitem[{{Knight} {et~al.}(2017){Knight}, {Protopapa}, {Kelley}, {Farnham},
  {Bauer}, {Bodewits}, {Feaga}, \& {Sunshine}}]{Knight2017}
{Knight}, M.~M., {Protopapa}, S., {Kelley}, M. S.~P., {et~al.} 2017, ApJL, 851,
  L31, \dodoi{10.3847/2041-8213/aa9d81}

\bibitem[{{Laughlin} \& {Batygin}(2017)}]{Laughlin2017}
{Laughlin}, G., \& {Batygin}, K. 2017, Research Notes of the American
  Astronomical Society, 1, 43, \dodoi{10.3847/2515-5172/aaa02b}

\bibitem[{{Levine} {et~al.}(2021){Levine}, {Cabot}, {Seligman}, \&
  {Laughlin}}]{Levine2021}
{Levine}, W.~G., {Cabot}, S. H.~C., {Seligman}, D., \& {Laughlin}, G. 2021,
  \apj, 922, 39, \dodoi{10.3847/1538-4357/ac1fe6}

\bibitem[{{Levine} \& {Laughlin}(2021)}]{Levine2021_h2}
{Levine}, W.~G., \& {Laughlin}, G. 2021, \apj, 912, 3,
  \dodoi{10.3847/1538-4357/abec85}

\bibitem[{{Lin} {et~al.}(2020){Lin}, {Lee}, {Gerdes}, {Adams}, {Becker},
  {Napier}, \& {Markwardt}}]{Lin2020}
{Lin}, H.~W., {Lee}, C.-H., {Gerdes}, D.~W., {et~al.} 2020, \apjl, 889, L30,
  \dodoi{10.3847/2041-8213/ab6bd9}

\bibitem[{{Loeb}(2018{\natexlab{a}})}]{Loeb2018a}
{Loeb}, A. 2018{\natexlab{a}}, Scientific American

\bibitem[{{Loeb}(2018{\natexlab{b}})}]{Loeb2018b}
---. 2018{\natexlab{b}}, Scientific American

\bibitem[{{Loeb}(2018{\natexlab{c}})}]{Loeb2018c}
---. 2018{\natexlab{c}}, Scientific American

\bibitem[{{Loeb}(2021{\natexlab{a}})}]{Loeb2021a}
---. 2021{\natexlab{a}}, (Houghton-Mifflin-Harcourt, New York

\bibitem[{{Loeb}(2021{\natexlab{b}})}]{Loeb2021b}
---. 2021{\natexlab{b}}, arXiv e-prints, arXiv:2110.15213.
\newblock \doarXiv{2110.15213}

\bibitem[{Luu {et~al.}(2020)Luu, Flekk{\o}y, \& Toussaint}]{luu2020oumuamua}
Luu, J.~X., Flekk{\o}y, E.~G., \& Toussaint, R. 2020, \apjl, 900, L22

\bibitem[{Lynch(2005)}]{lynch2005infrared}
Lynch, D. 2005, The infrared spectral signature of water ice in the vacuum
  cryogenic AI\&T environment, Tech. rep., AEROSPACE CORP EL SEGUNDO CA

\bibitem[{{Mamajek}(2017)}]{mamajek2017}
{Mamajek}, E. 2017, Research Notes of the American Astronomical Society, 1, 21,
  \dodoi{10.3847/2515-5172/aa9bdc}

\bibitem[{{Manzini} {et~al.}(2020){Manzini}, {Oldani}, {Ochner}, \&
  {Bedin}}]{Manzini2020}
{Manzini}, F., {Oldani}, V., {Ochner}, P., \& {Bedin}, L.~R. 2020, \mnras, 495,
  L92, \dodoi{10.1093/mnrasl/slaa061}

\bibitem[{{Mashchenko}(2019)}]{Mashchenko2019}
{Mashchenko}, S. 2019, MNRAS, 489, 3003, \dodoi{10.1093/mnras/stz2380}

\bibitem[{{Masiero}(2017)}]{masiero2017spectrum}
{Masiero}, J. 2017, arXiv e-prints, arXiv:1710.09977.
\newblock \doarXiv{1710.09977}

\bibitem[{{McKay} {et~al.}(2020){McKay}, {Cochran}, {Dello Russo}, \&
  {DiSanti}}]{McKay2020}
{McKay}, A.~J., {Cochran}, A.~L., {Dello Russo}, N., \& {DiSanti}, M.~A. 2020,
  \apjl, 889, L10, \dodoi{10.3847/2041-8213/ab64ed}

\bibitem[{{McNeill} {et~al.}(2018){McNeill}, {Trilling}, \&
  {Mommert}}]{McNeill2018}
{McNeill}, A., {Trilling}, D.~E., \& {Mommert}, M. 2018, ApJL, 857, L1,
  \dodoi{10.3847/2041-8213/aab9ab}

\bibitem[{{Meech} {et~al.}(2019){Meech}, {Castillo-Rogez}, {Hainaut}, {Lazio},
  \& {Raymond}}]{Meech2019whitepaper}
{Meech}, K., {Castillo-Rogez}, J., {Hainaut}, O., {Lazio}, J., \& {Raymond}, S.
  2019, \baas, 51, 552

\bibitem[{{Meech} {et~al.}(2021){Meech}, {Castillo-Rogez}, {Bufanda}, {Buie},
  {Hainaut}, {Ishii}, {Keane}, {Kleyna}, {Li}, {Raymond}, {Wainscoat}, {Weryk},
  \& {Yang}}]{Meech2021}
{Meech}, K., {Castillo-Rogez}, J., {Bufanda}, E., {et~al.} 2021, in Bulletin of
  the American Astronomical Society, Vol.~53, 282,
  \dodoi{10.3847/25c2cfeb.ea404475}

\bibitem[{{Meech} {et~al.}(2017){Meech}, {Weryk}, {Micheli}, {Kleyna},
  {Hainaut}, {Jedicke}, {Wainscoat}, {Chambers}, {Keane}, {Petric}, {Denneau},
  {Magnier}, {Berger}, {Huber}, {Flewelling}, {Waters}, {Schunova-Lilly}, \&
  {Chastel}}]{Meech2017}
{Meech}, K.~J., {Weryk}, R., {Micheli}, M., {et~al.} 2017, Nature, 552, 378,
  \dodoi{10.1038/nature25020}

\bibitem[{{Mencos} \& {Krim}(2018)}]{2018MNRAS.476.5432M}
{Mencos}, A., \& {Krim}, L. 2018, \mnras, 476, 5432,
  \dodoi{10.1093/mnras/sty609}

\bibitem[{{Metzger} {et~al.}(2019){Metzger}, {Berger}, {Grindlay}, {Gezari},
  {Iveziv}, {Jencson}, {Kasliwal}, {Kutyrev}, {Macleod}, {Melnick}, {Purcell},
  {Rieke}, {Shen}, {Tanvi}, \& {Vasey}}]{2019astro2020T.306M}
{Metzger}, B., {Berger}, E., {Grindlay}, J., {et~al.} 2019, Astro2020: Decadal
  Survey on Astronomy and Astrophysics, 2020, 306.
\newblock \doarXiv{1903.05736}

\bibitem[{{Micheli} {et~al.}(2018){Micheli}, {Farnocchia}, {Meech}, {Buie},
  {Hainaut}, {Prialnik}, {Sch{\"o}rghofer}, {Weaver}, {Chodas}, \&
  {Kleyna}}]{Micheli2018}
{Micheli}, M., {Farnocchia}, D., {Meech}, K.~J., {et~al.} 2018, Nature, 559,
  223, \dodoi{10.1038/s41586-018-0254-4}

\bibitem[{Mongelli(2019)}]{mongelli2019albedo}
Mongelli, G.~F. 2019, International Journal of Chemistry, Mathematics And
  Physics, 3, 18

\bibitem[{{Moore} {et~al.}(2021{\natexlab{a}}){Moore}, {Castillo-Rogez},
  {Meech}, {Courville}, {Donitz}, {Ferguson}, {Llera}, \&
  {French}}]{Moore2021whitepaper}
{Moore}, K., {Castillo-Rogez}, J., {Meech}, K.~J., {et~al.} 2021{\natexlab{a}},
  in Bulletin of the American Astronomical Society, Vol.~53, 481,
  \dodoi{10.3847/25c2cfeb.1d58e5af}

\bibitem[{{Moore} {et~al.}(2021{\natexlab{b}}){Moore}, {Courville}, {Ferguson},
  {Schoenfeld}, {Llera}, {Agrawal}, {Brack}, {Buhler}, {Connour}, {Czaplinski},
  {DeLuca}, {Deutsch}, {Hammond}, {Kuettel}, {Marusiak}, {Nerozzi}, {Stuart},
  {Tarnas}, {Thelen}, {Castillo-Rogez}, {Smythe}, {Landau}, {Mitchell}, \&
  {Budney}}]{Moore2021}
{Moore}, K., {Courville}, S., {Ferguson}, S., {et~al.} 2021{\natexlab{b}},
  \planss, 197, 105137, \dodoi{10.1016/j.pss.2020.105137}

\bibitem[{{Morbidelli} {et~al.}(2005){Morbidelli}, {Levison}, {Tsiganis}, \&
  {Gomes}}]{Morbidelli2005}
{Morbidelli}, A., {Levison}, H.~F., {Tsiganis}, K., \& {Gomes}, R. 2005, \nat,
  435, 462, \dodoi{10.1038/nature03540}

\bibitem[{{Moro-Mart{\'{\i}}n}(2018)}]{moro2018}
{Moro-Mart{\'{\i}}n}, A. 2018, \apj, 866, 131, \dodoi{10.3847/1538-4357/aadf34}

\bibitem[{{Moro-Mart{\'{\i}}n}(2019)}]{moro2019a}
---. 2019, \aj, 157, 86, \dodoi{10.3847/1538-3881/aafda6}

\bibitem[{Moro-Mart{\'\i}n(2019)}]{moro2019fractal}
Moro-Mart{\'\i}n, A. 2019, \apjl, 872, L32

\bibitem[{{Moro-Mart{\'\i}n} \& {Norman}(2022)}]{Moro2022}
{Moro-Mart{\'\i}n}, A., \& {Norman}, C. 2022, \apj, 924, 96,
  \dodoi{10.3847/1538-4357/ac32cc}

\bibitem[{{Moro-Mart{\'{\i}}n} {et~al.}(2009){Moro-Mart{\'{\i}}n}, {Turner}, \&
  {Loeb}}]{Moro2009}
{Moro-Mart{\'{\i}}n}, A., {Turner}, E.~L., \& {Loeb}, A. 2009, \apj, 704, 733,
  \dodoi{10.1088/0004-637X/704/1/733}

\bibitem[{{Okuzumi} {et~al.}(2012){Okuzumi}, {Tanaka}, {Kobayashi}, \&
  {Wada}}]{Okuzumi2012}
{Okuzumi}, S., {Tanaka}, H., {Kobayashi}, H., \& {Wada}, K. 2012, \apj, 752,
  106, \dodoi{10.1088/0004-637X/752/2/106}

\bibitem[{{Opitom} {et~al.}(2019){Opitom}, {Fitzsimmons}, {Jehin}, {Moulane},
  {Hainaut}, {Meech}, {Yang}, {Snodgrass}, {Micheli}, {Keane}, {Benkhaldoun},
  \& {Kleyna}}]{Opitom:2019-borisov}
{Opitom}, C., {Fitzsimmons}, A., {Jehin}, E., {et~al.} 2019, \aap, 631, L8,
  \dodoi{10.1051/0004-6361/201936959}

\bibitem[{{'Oumuamua ISSI Team} {et~al.}(2019){'Oumuamua ISSI Team},
  {Bannister}, {Bhandare}, {Dybczy{\'n}ski}, {Fitzsimmons},
  {Guilbert-Lepoutre}, {Jedicke}, {Knight}, {Meech}, {McNeill}, {Pfalzner},
  {Raymond}, {Snodgrass}, {Trilling}, \& {Ye}}]{ISSI2019}
{'Oumuamua ISSI Team}, {Bannister}, M.~T., {Bhandare}, A., {et~al.} 2019,
  Nature Astronomy, 3, 594, \dodoi{10.1038/s41550-019-0816-x}

\bibitem[{{Pau S{\'a}nchez} {et~al.}(2021){Pau S{\'a}nchez}, {Morante},
  {Hermosin}, {Ranuschio}, {Estalella}, {Viera}, {Centuori}, {Jones},
  {Snodgrass}, {Chantal Levasseur-Regourd}, \& {Tubiana}}]{Sanchez2021}
{Pau S{\'a}nchez}, J., {Morante}, D., {Hermosin}, P., {et~al.} 2021, arXiv
  e-prints, arXiv:2107.12999.
\newblock \doarXiv{2107.12999}

\bibitem[{Pfalzner \& Bannister(2019)}]{Pfalzner2019}
Pfalzner, S., \& Bannister, M.~T. 2019, \apjl, 874, L34

\bibitem[{{Pfalzner} {et~al.}(2021){Pfalzner}, {Paterson}, {Bannister}, \&
  {Portegies Zwart}}]{Pfalzner2021}
{Pfalzner}, S., {Paterson}, D., {Bannister}, M.~T., \& {Portegies Zwart}, S.
  2021, \apj, 921, 168, \dodoi{10.3847/1538-4357/ac0c10}

\bibitem[{{Rafikov}(2018)}]{Rafikov2018}
{Rafikov}, R.~R. 2018, \apjl, 867, L17, \dodoi{10.3847/2041-8213/aae977}

\bibitem[{{Raymond} {et~al.}(2018{\natexlab{a}}){Raymond}, {Armitage}, \&
  {Veras}}]{Raymond2018b}
{Raymond}, S.~N., {Armitage}, P.~J., \& {Veras}, D. 2018{\natexlab{a}}, \apjl,
  856, L7, \dodoi{10.3847/2041-8213/aab4f6}

\bibitem[{{Raymond} {et~al.}(2018{\natexlab{b}}){Raymond}, {Armitage}, {Veras},
  {Quintana}, \& {Barclay}}]{Raymond2017}
{Raymond}, S.~N., {Armitage}, P.~J., {Veras}, D., {Quintana}, E.~V., \&
  {Barclay}, T. 2018{\natexlab{b}}, \mnras, 476, 3031,
  \dodoi{10.1093/mnras/sty468}

\bibitem[{Richardson {et~al.}(2004)Richardson, Melosh, \&
  Greenberg}]{doi:10.1126/science.1104731}
Richardson, J.~E., Melosh, H.~J., \& Greenberg, R. 2004, Science, 306, 1526,
  \dodoi{10.1126/science.1104731}

\bibitem[{Sasaki {et~al.}(2001)Sasaki, Nakamura, Hamabe, Kurahashi, \&
  Hiroi}]{sasaki2001production}
Sasaki, S., Nakamura, K., Hamabe, Y., Kurahashi, E., \& Hiroi, T. 2001, Nature,
  410, 555

\bibitem[{{Sekanina}(2019{\natexlab{a}})}]{Sekanina2019b}
{Sekanina}, Z. 2019{\natexlab{a}}, arXiv e-prints.
\newblock \doarXiv{1901.08704}

\bibitem[{{Sekanina}(2019{\natexlab{b}})}]{Sekanina2019}
---. 2019{\natexlab{b}}, arXiv e-prints, arXiv:1901.08704.
\newblock \doarXiv{1901.08704}

\bibitem[{{Seligman} \& {Laughlin}(2018)}]{Seligman2018}
{Seligman}, D., \& {Laughlin}, G. 2018, \aj, 155, 217,
  \dodoi{10.3847/1538-3881/aabd37}

\bibitem[{{Seligman} \& {Laughlin}(2020{\natexlab{a}})}]{seligman2020}
---. 2020{\natexlab{a}}, \apjl, 896, L8, \dodoi{10.3847/2041-8213/ab963f}

\bibitem[{{Seligman} \& {Laughlin}(2020{\natexlab{b}})}]{2020ApJ...896L...8S}
---. 2020{\natexlab{b}}, \apjl, 896, L8, \dodoi{10.3847/2041-8213/ab963f}

\bibitem[{{Seligman} {et~al.}(2019){Seligman}, {Laughlin}, \&
  {Batygin}}]{seligman2019acceleration}
{Seligman}, D., {Laughlin}, G., \& {Batygin}, K. 2019, \apjl, 876, L26

\bibitem[{{Seligman} {et~al.}(2021){Seligman}, {Levine}, {Cabot}, {Laughlin},
  \& {Meech}}]{Seligman2021}
{Seligman}, D.~Z., {Levine}, W.~G., {Cabot}, S. H.~C., {Laughlin}, G., \&
  {Meech}, K. 2021, \apj, 920, 28, \dodoi{10.3847/1538-4357/ac1594}

\bibitem[{{Shen} {et~al.}(2019){Shen}, {Anderson}, {Berger}, {Brandt}, {De
  Rosa}, {Fan}, {Ferrarese}, {Gezari}, {Graham}, {Greene}, {Grier}, {Grindlay},
  {Haggard}, {Hall}, {Ho}, {Ibarra Medel}, {Ilic}, {Ivezic}, {Jencson},
  {Jiang}, {Juneau}, {Kasliwal}, {Kollmeier}, {Kutyrev}, {I-Hsiu Li}, {Liu},
  {Liu}, {MacLeod}, {Melnick}, {Metzger}, {Myers}, {O'Dea}, {Petric},
  {Popovi{\'c}}, {Prakash}, {Purcell}, {Richards}, {Rieke}, {Tanvir},
  {Trakhtenbrot}, {Wood-Vasey}, {Xue}, \& {Yang}}]{2019BAAS...51c.274S}
{Shen}, Y., {Anderson}, S., {Berger}, E., {et~al.} 2019, \baas, 51, 274.
\newblock \doarXiv{1903.04533}

\bibitem[{{Siraj} \& {Loeb}(2019)}]{2019arXiv190407224S}
{Siraj}, A., \& {Loeb}, A. 2019, arXiv e-prints, arXiv:1904.07224.
\newblock \doarXiv{1904.07224}

\bibitem[{{Siraj} \& {Loeb}(2022)}]{Siraj2022}
---. 2022, \na, 92, 101730, \dodoi{10.1016/j.newast.2021.101730}

\bibitem[{Solontoi {et~al.}(2011)Solontoi, Ivezi{\'c}, \&
  Jones}]{solontoi2011comet}
Solontoi, M., Ivezi{\'c}, {\v{Z}}., \& Jones, L. 2011, in American Astronomical
  Society Meeting Abstracts\# 217, Vol. 217, 252--11

\bibitem[{Stern {et~al.}(2019)Stern, Weaver, Spencer, Olkin, Gladstone, Grundy,
  Moore, Cruikshank, Elliott, McKinnon, Parker, Verbiscer, Young, Aguilar,
  Albers, Andert, Andrews, Bagenal, Banks, Bauer, Bauman, Bechtold,
  Beddingfield, Behrooz, Beisser, Benecchi, Bernardoni, Beyer, Bhaskaran,
  Bierson, Binzel, Birath, Bird, Boone, Bowman, Bray, Britt, Brown, Buckley,
  Buie, Buratti, Burke, Bushman, Carcich, Chaikin, Chavez, Cheng, Colwell,
  Conard, Conner, Conrad, Cook, Cooper, Custodio, Ore, Deboy, Dharmavaram,
  Dhingra, Dunn, Earle, Egan, Eisig, El-Maarry, Engelbrecht, Enke, Ercol,
  Fattig, Ferrell, Finley, Firer, Fischetti, Folkner, Fosbury, Fountain,
  Freeze, Gabasova, Glaze, Green, Griffith, Guo, Hahn, Hals, Hamilton,
  Hamilton, Hanley, Harch, Harmon, Hart, Hayes, Hersman, Hill, Hill,
  Hofgartner, Holdridge, Horányi, Hosadurga, Howard, Howett, Jaskulek,
  Jennings, Jensen, Jones, Kang, Katz, Kaufmann, Kavelaars, Keane, Keleher,
  Kinczyk, Kochte, Kollmann, Krimigis, Kruizinga, Kusnierkiewicz, Lahr, Lauer,
  Lawrence, Lee, Lessac-Chenen, Linscott, Lisse, Lunsford, Mages, Mallder,
  Martin, May, McComas, McNutt, Mehoke, Mehoke, Nelson, Nguyen, Núñez,
  Ocampo, Owen, Oxton, Parker, Pätzold, Pelgrift, Pelletier, Pineau, Piquette,
  Porter, Protopapa, Quirico, Redfern, Regiec, Reitsema, Reuter, Richardson,
  Riedel, Ritterbush, Robbins, Rodgers, Rogers, Rose, Rosendall, Runyon,
  Ryschkewitsch, Saina, Salinas, Schenk, Scherrer, Schlei, Schmitt, Schultz,
  Schurr, Scipioni, Sepan, Shelton, Showalter, Simon, Singer, Stahlheber,
  Stanbridge, Stansberry, Steffl, Strobel, Stothoff, Stryk, Stuart, Summers,
  Tapley, Taylor, Taylor, Tedford, Throop, Turner, Umurhan, Eck, Velez,
  Versteeg, Vincent, Webbert, Weidner, Weigle, Wendel, White, Whittenburg,
  Williams, Williams, Williams, Winters, Zangari, \&
  Zurbuchen}]{doi:10.1126/science.aaw9771}
Stern, S.~A., Weaver, H.~A., Spencer, J.~R., {et~al.} 2019, Science, 364,
  eaaw9771, \dodoi{10.1126/science.aaw9771}

\bibitem[{{Tanvir} {et~al.}(2019){Tanvir}, {Grindlay}, {Berger}, {Metzger},
  {Gezari}, {Ivezic}, {Jencson}, {Kasliwal}, {Kutyrev}, {Macleod}, {Melnick},
  {Purcell}, {Rieke}, {Shen}, \& {Vasey}}]{2019BAAS...51c.472T}
{Tanvir}, N., {Grindlay}, J., {Berger}, E., {et~al.} 2019, \baas, 51, 472.
\newblock \doarXiv{1903.07835}

\bibitem[{{Trilling} {et~al.}(2017){Trilling}, {Robinson}, {Roegge}, {Chand
  ler}, {Smith}, {Loeffler}, {Trujillo}, {Navarro-Meza}, \&
  {Glaspie}}]{Trilling2017}
{Trilling}, D.~E., {Robinson}, T., {Roegge}, A., {et~al.} 2017, \apjl, 850,
  L38, \dodoi{10.3847/2041-8213/aa9989}

\bibitem[{{Trilling} {et~al.}(2018){Trilling}, {Mommert}, {Hora}, {Farnocchia},
  {Chodas}, {Giorgini}, {Smith}, {Carey}, {Lisse}, {Werner}, {McNeill},
  {Chesley}, {Emery}, {Fazio}, {Fernandez}, {Harris}, {Marengo}, {Mueller},
  {Roegge}, {Smith}, {Weaver}, {Meech}, \& {Micheli}}]{Trilling2018}
{Trilling}, D.~E., {Mommert}, M., {Hora}, J.~L., {et~al.} 2018, \aj, 156, 261,
  \dodoi{10.3847/1538-3881/aae88f}

\bibitem[{{Tsiganis} {et~al.}(2005){Tsiganis}, {Gomes}, {Morbidelli}, \&
  {Levison}}]{Tsiganis2005}
{Tsiganis}, K., {Gomes}, R., {Morbidelli}, A., \& {Levison}, H.~F. 2005, \nat,
  435, 459, \dodoi{10.1038/nature03539}

\bibitem[{{Vere{\v{s}}} \& {Chesley}(2017{\natexlab{a}})}]{Veres2017}
{Vere{\v{s}}}, P., \& {Chesley}, S.~R. 2017{\natexlab{a}}, \aj, 154, 13,
  \dodoi{10.3847/1538-3881/aa73d0}

\bibitem[{{Vere{\v{s}}} \& {Chesley}(2017{\natexlab{b}})}]{veres2017b}
---. 2017{\natexlab{b}}, \aj, 154, 12, \dodoi{10.3847/1538-3881/aa73d1}

\bibitem[{{Viviano-Beck} {et~al.}(2014){Viviano-Beck}, {Seelos}, {Murchie},
  {Kahn}, {Seelos}, {Taylor}, {Taylor}, {Ehlmann}, {Wisemann}, {Mustard}, \&
  {Morgan}}]{2014JGRE..119.1403V}
{Viviano-Beck}, C.~E., {Seelos}, F.~P., {Murchie}, S.~L., {et~al.} 2014,
  Journal of Geophysical Research (Planets), 119, 1403,
  \dodoi{10.1002/2014JE004627}

\bibitem[{{Xing} {et~al.}(2020){Xing}, {Bodewits}, {Noonan}, \&
  {Bannister}}]{Xing2020}
{Xing}, Z., {Bodewits}, D., {Noonan}, J., \& {Bannister}, M.~T. 2020, \apjl,
  893, L48, \dodoi{10.3847/2041-8213/ab86be}

\bibitem[{Yang {et~al.}(2021)Yang, Li, Cordiner, Chang, Hainaut, Williams,
  Meech, Keane, \& Villard}]{yang2021}
Yang, B., Li, A., Cordiner, M.~A., {et~al.} 2021, Nature Astronomy,
  \dodoi{10.1038/s41550-021-01336-w}

\bibitem[{{Ye} {et~al.}(2020){Ye}, {Kelley}, {Bolin}, {Bodewits}, {Farnocchia},
  {Masci}, {Meech}, {Micheli}, {Weryk}, {Bellm}, {Christensen}, {Dekany},
  {Delacroix}, {Graham}, {Kulkarni}, {Laher}, {Rusholme}, \& {Smith}}]{Ye:2019}
{Ye}, Q., {Kelley}, M. S.~P., {Bolin}, B.~T., {et~al.} 2020, \aj, 159, 77,
  \dodoi{10.3847/1538-3881/ab659b}

\bibitem[{{Zhang} {et~al.}(2020){Zhang}, {Ye}, \& {Kolokolova}}]{Zhang2020ATel}
{Zhang}, Q., {Ye}, Q., \& {Kolokolova}, L. 2020, The Astronomer's Telegram,
  13618, 1

\bibitem[{{Zheng} {et~al.}(2009){Zheng}, {Jewitt}, \&
  {Kaiser}}]{2009ApJS..181...53Z}
{Zheng}, W., {Jewitt}, D., \& {Kaiser}, R.~I. 2009, \apjs, 181, 53,
  \dodoi{10.1088/0067-0049/181/1/53}

\bibitem[{{Zwart} {et~al.}(2018){Zwart}, {Torres}, {Pelupessy}, {B{\'e}dorf},
  \& {Cai}}]{Zwart2018}
{Zwart}, P., S., {Torres}, S., {Pelupessy}, I., {B{\'e}dorf}, J., \& {Cai},
  M.~X. 2018, \mnras, 479, L17, \dodoi{10.1093/mnrasl/sly088}

\end{thebibliography}
\bibliographystyle{aasjournal}



\end{document}